\documentclass{article}

\usepackage{xr-hyper}
\makeatletter
    \newcommand*{\addFileDependency}[1]{
      \typeout{(#1)}
      \@addtofilelist{#1}
      \IfFileExists{#1}{}{\typeout{No file #1.}}
    }
\makeatother
\newcommand*{\myexternaldocument}[3][supp]{%
    \externaldocument[#1]{#2/#3}%
    \addFileDependency{#3.tex}%
    \addFileDependency{#2/#3.aux}%
}
\myexternaldocument[supp]{build}{appendix}

\usepackage[final,nonatbib]{neurips_2022}

\usepackage[utf8]{inputenc} 
\usepackage[T1]{fontenc}    
\usepackage{hyperref}       
\usepackage{url}            
\usepackage{booktabs}       
\usepackage{amsfonts}       
\usepackage{nicefrac}       
\usepackage{microtype}      

\usepackage{amsmath}

\DeclareMathOperator*{\argmin}{arg\,min}

\usepackage[dvipsnames]{xcolor}
\usepackage{tikz}
\usetikzlibrary{backgrounds}
\usetikzlibrary{arrows,shapes}
\usetikzlibrary{tikzmark}
\usetikzlibrary{calc}

\usepackage{amsmath}
\usepackage{amsthm}
\usepackage{amssymb}
\usepackage{mathtools, nccmath}
\usepackage{wrapfig}
\usepackage{comment}

\usepackage{blindtext}

\usepackage{graphicx,import}

\usepackage{xspace}

\usepackage{array}
\usepackage{ragged2e}
\newcolumntype{P}[1]{>{\RaggedRight\hspace{0pt}}p{#1}}
\newcolumntype{X}[1]{>{\RaggedRight\hspace*{0pt}}p{#1}}

\usepackage{tcolorbox}

\usepackage{tikz}
\usetikzlibrary{arrows,shapes,positioning,shadows,trees,mindmap}
\usepackage[edges]{forest}
\usetikzlibrary{arrows.meta}
\colorlet{linecol}{black!75}
\usepackage{xkcdcolors} 

\usepackage{tikz}
\usetikzlibrary{backgrounds}
\usetikzlibrary{arrows,shapes}
\usetikzlibrary{tikzmark}
\usetikzlibrary{calc}
\newcommand{\highlight}[2]{\colorbox{#1!10}{$\displaystyle #2$}}

\colorlet{mhpurple}{Plum!80}

\renewcommand{\highlight}[2]{\colorbox{#1!10}{#2}}

\newlength\myindent
\usepackage{algorithm}
\usepackage{algorithmicx}
\usepackage{algpseudocode}
\usepackage{bbold}
\usepackage{bm}
\usepackage[labelfont=bf]{caption}
\usepackage{subcaption}

\usepackage{longtable}
\usepackage{supertabular,booktabs}

\usepackage{graphicx}
\usepackage{multicol}
\usepackage{svg}

\usepackage{biblatex}
\newcommand{\beginsupplement}{%
    \setcounter{table}{0}
    \renewcommand{\thetable}{S\arabic{table}}%
    \setcounter{figure}{0}
    \renewcommand{\thefigure}{S\arabic{figure}}%
    \setcounter{equation}{0}
    \renewcommand{\theequation}{S\arabic{equation}}%
    \setcounter{algorithm}{0}
    \renewcommand{\thealgorithm}{S\arabic{algorithm}}
}

\DeclareUnicodeCharacter{2212}{-}

\newcommand{\bs}[1]{\boldsymbol{#1}}

\title{
    Aligning individual brains\\
    with Fused Unbalanced Gromov-Wasserstein
}

\author{
  Alexis Thual\footnotemark[1]{}\\
  Cognitive Neuroimaging Unit, INSERM, CEA, CNRS, NeuroSpin center, Gif sur Yvette, France\\
  Mind, Inria Paris-Saclay, Palaiseau, France\\
  Inserm, Collège de France, Paris, France\\
  \texttt{alexis.thual@cea.fr}
  \And
  Huy Tran\footnotemark[1]{}\\
  CMAP, Ecole Polytechnique, Palaiseau, France\\
  Université Bretagne-Sud, CNRS, IRISA, Vannes, France\\
  \texttt{quang-huy.tran@univ-ubs.fr}
  \And
  Tatiana Zemskova\\
  Mind, Inria Paris-Saclay, Palaiseau, France\\
  \texttt{tatiana.zemskova@polytechnique.edu}
  \AND
  Nicolas Courty\\
  Université Bretagne-Sud, CNRS, IRISA, Vannes, France\\
  \texttt{ncourty@irisa.fr}
  \And
  Rémi Flamary\\
  CMAP, Ecole Polytechnique, Palaiseau, France\\
  \texttt{remi.flamary@polytechnique.edu}
  \And
  Stanislas Dehaene\\
  Cognitive Neuroimaging Unit, INSERM, CEA, CNRS, NeuroSpin center, Gif sur Yvette, France\\
  Inserm, Collège de France, Paris, France\\
  \texttt{stanislas.dehaene@cea.fr}
  \And
  Bertrand Thirion\\
  Mind, Inria Paris-Saclay, Palaiseau, France\\
  \texttt{bertrand.thirion@inria.fr}
}

\begin{document}

\renewcommand*{\thefootnote}{\fnsymbol{footnote}}

\maketitle

\footnotetext[1]{These authors contributed equally.}

\newpage

\renewcommand*{\thefootnote}{\arabic{footnote}}

\begin{abstract}
    Individual brains vary in both anatomy and functional organization, even within a given species. 
    Inter-individual variability is a major impediment when trying to draw generalizable conclusions from neuroimaging data collected on groups of subjects.
    Current co-registration procedures rely on limited data, and thus lead to very coarse inter-subject alignments. 
    In this work, we present a novel method for inter-subject alignment based on Optimal Transport, denoted as
    Fused Unbalanced Gromov Wasserstein (FUGW). The method aligns cortical surfaces based on the similarity of their functional signatures in response to a variety of stimulation settings, while penalizing large deformations of 
    individual topographic organization.
    %
    We demonstrate that FUGW is well-suited for whole-brain landmark-free alignment. The unbalanced feature allows to deal with the fact that functional areas
    vary in size across subjects. Our results show that FUGW alignment significantly increases between-subject correlation of activity for independent functional data, and leads to more precise mapping at the group level.
\end{abstract}

\section{Introduction}
\label{sec:introduction}

\begin{figure}[ht]
    \centering
    \includegraphics[width=\columnwidth]{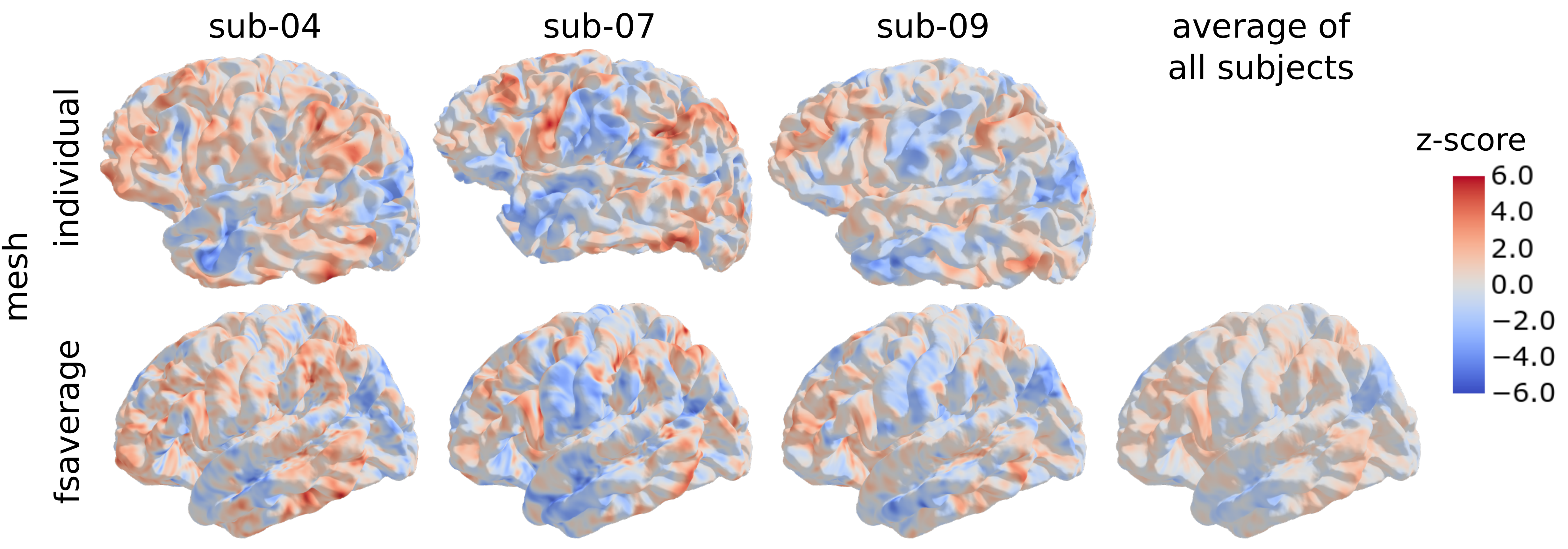}
    \caption{
        \textbf{High variability in human anatomies and functional MRI responses across subjects}
        In this experiment contrasting areas of the brain
        which respond to mathematical tasks against other that
        don't, we observe great variability in locations and strength of brain activations across subjects (row 1).
        The classical approach consists in wrapping this data
        to a common surface template (row 2), where they can be averaged, often resulting in
        loss of individual details and detection power. These images were generated using Nilearn software \cite{abraham_machine_2014}.
    }
    \label{fig:intro}
\end{figure}

The availability of millimeter or sub-millimeter anatomical or
functional brain images has opened new horizons to
neuroscience, namely that of mapping cognition in the human brain and
detecting markers of diseases.
Yet this endeavour has stumbled on the roadblock of inter-individual
variability: while the overall organization of the human brain is largely invariant, two different brains (even from monozygotic twins \cite{pizzigali2020})
may differ at the scale of centimeters in shape, folding pattern, and functional responses.
%
The problem is further complicated by the fact that functional images
are noisy, due to imaging limitations and behavioral
differences across individuals that cannot be easily overcome.
The status quo of the field is thus to rely on anatomy-based inter-individual alignment that approximately matches the outline of the brain \cite{ants,} as well as its large-scale cortical folding patterns \cite{fs_reconall,fischl_freesurfer_2012}.
Existing algorithms thus coarsely match anatomical features with diffeomorphic transformations, by warping individual data to a simplified template brain.
Such methods lose much of the original individual detail and blur the functional information that can be measured in brain regions (see Figure \ref{fig:intro}).
%

In order to improve upon the current situation, a number of challenges have to be addressed:
\textit{(i)} There exists no template brain with functional information, which by construction renders any cortical matching method blind to function. This is unfortunate, since functional information is arguably the most accessible marker to identify cortical regions and their boundaries \cite{Glasser2016-ha}.
\textit{(ii)} When comparing two brains -- coming from individuals or from a template -- it is unclear what regularity should be imposed on the matching \cite{vanessen2012}. While it is traditional in medical imaging to impose diffeomorphicity \cite{ants}, such a constrain does not match the frequent observation that brain regions vary across individuals in their fine-grained functional organization \cite{Glasser2016-ha,schneider2019}.
%
\textit{(iii)} Beyond the problem of aligning human brains, it is an even greater challenge to systematically compare functional brain organization in two different species, such as humans and macaques \cite{neubert_comparison_2014,mars_whole_2018,xu_cross-species_2020,eichert_cross-species_2020,}. 
Such inter-species comparisons introduce a more extreme form of variability in the correspondence model. 

%
%


\paragraph{Related work}
Several attempts have been made to constrain the brain alignment process by using functional information.
The first one consists in introducing functional maps into the diffeomorphic framework and search for a smooth transformation that matches functional information \cite{sabuncu_function-based_2010,yeo_spherical_2010,robinson_msm_2014}, the most popular framework being arguably Multimodal Surface Matching (MSM) \cite{robinson_msm_2014,Glasser2016-ha}.

A second family of less constrained functional alignment approaches have been proposed, based on heuristics, by matching information in small, possibly overlapping, cortical patches \cite{haxby_common_2011,Tavor2016-rl,bazeille_empirical_2021}. 
This popular framework has been called \emph{hyperalignment} \cite{haxby_common_2011,guntupalli_model_2016}, or \emph{shared response models} \cite{Chen2015}. 
Yet these approaches lack a principled framework and cannot be considered to solve the matching problem at scale. Neither do they allow to estimate %
a group-level template properly \cite{alwasity2020}.  
%

An alternative functional alignment framework has followed another path \cite{gramfort2015}, considering functional signal as a three-dimensional distribution, and minimizing the transport cost. However, this framework imposes unnatural constraints of non-negativity of the signal and 
%
only works for one-dimensional contrasts, so that it cannot be used to learn multi-dimensional anatomo-functional structures.
An important limitation of the latter two families of methods is that they operate on a fixed spatial context (mesh or voxel grid), and thus cannot be used on heterogeneous meshes such as between two individual human anatomies or, worse, between a monkey brain and a human brain.
%

\paragraph{Contributions}

%
Following \cite{bazeille_local_2019}, we use the Wasserstein distance between source and target functional signals -- consisting of contrast maps acquired with fMRI -- to compute brain alignments.
%
We contribute two notable extensions of this framework:
\textit{(i)} a Gromov-Wasserstein (GW) term to preserve global anatomical structure -- this term introduces an anatomical penalization against improbably distant anatomical matches, yet without imposing diffeomorphic regularity --
as well as \textit{(ii)} an unbalanced correspondence that allows mappings from one brain to another to be incomplete, for instance because some functional areas are larger in some individuals than in others, or may simply be absent.
%
%
%
%
We show that this approach successfully addresses the challenging case of different cortical meshes, and that derived brain activity templates are sharper than those obtained with standard anatomical alignment approaches.

%

\section{Methods}
\label{sec:methods}

Optimal Transport yields a natural framework to address the alignment problem, as it seeks to derive a plan -- a \textit{coupling} -- that can be seen as a soft assignment matrix between cortical areas of a source and target individual.
As discussed previously, there is a need for a functional alignment method that respects the rich geometric structure of the anatomical features, hence the Wasserstein distance alone is not sufficient.
By construction, the GW distance \cite{memoli_use_2007,memoli_gromovwasserstein_2011} can help preserve the global geometry underlying the signal. The more recent fused GW distance \cite{vayer_fused_2018} goes one step further by making it possible to integrate functional data simultaneously with anatomical information.

\subsection{Fused Unbalanced Gromov-Wasserstein}

We leverage \cite{vayer_fused_2018,sejourne_unbalanced_2021} to present a new objective function which interpolates between a loss preserving the global geometry of the underlying mesh structure and a loss aligning source and target features, while simultaneously allowing not to transport some parts of the source and target distributions. We provide an open-source solver that minimizes this loss\footnote{\href{https://github.com/alexisthual/fugw}{https://github.com/alexisthual/fugw} provides a PyTorch \cite{NEURIPS2019_9015} solver with a scikit-learn \cite{scikit-learn} compatible API}.


\paragraph{Formulation}

%

We denote $\bm{F^s} \in \mathbb R^{n, c}$ the matrix of features per vertex for the source subject.
In the proposed application, they correspond to $c$ functional activation maps, sampled on a mesh with $n$ vertices representing the source subject's cortical surface.
%
%
Let $\bm{D^s} \in \mathbb R^{n, n}_+$ be the matrix of pairwise geodesic distances\footnote{We compute geodesic distances using \href{https://github.com/the-virtual-brain/tvb-gdist}{https://github.com/the-virtual-brain/tvb-gdist}} between vertices of the source mesh.
Moreover, we assign the distribution ${\bm{w^s}} \in \mathbb R^{n}_+$ on the source vertices.
Comparably, we define $\bm{F^t} \in \mathbb R^{p, c}$, $\bm{D^t} \in \mathbb R^{p, p}_+$ and ${\bm{w^t}} \in \mathbb R^{p}_+$ for the target subject, whose individual anatomy is represented by a mesh comprising $p$ vertices.
Eventually, $\bm{w^s}$ and $\bm{w^t}$ set the transportable mass per vertex, which, without prior knowledge, we choose to be uniform for the source and target vertices respectively:
${\bm{w^s}} \triangleq (\frac{1}{n}, ..., \frac{1}{n})$,
${\bm{w^t}} \triangleq (\frac{1}{p}, ..., \frac{1}{p})$.

Given a tuple of hyper-parameters $\theta \triangleq (\rho, \alpha, \varepsilon)$,
where $\rho, \varepsilon \in \mathbb R_+$ and $\alpha \in [0,1]$,
for any coupling $\bm{P} \in \mathbb R^{n, p}$,
we define the fused unbalanced Gromov-Wasserstein loss as

\vspace*{1\baselineskip}
\begin{equation}
    \label{eq:fugw_loss}
    \begin{split}
    \text{L}_{\theta}(\bm{P}) \enspace \triangleq \enspace
    &(1 - \alpha) \enspace
    \tikzmarknode{wass}{
        \highlight{blue}{
            $\sum\limits_{\substack{0 \leq i < n\\0 \leq j < p}} || \bm{F^s}_i - \bm{F^t}_j||_2^2 \bm{P}_{i,j}$
        }
    }
    + \alpha \enspace
    \tikzmarknode{gw}{
        \highlight{teal}{
            $\sum\limits_{\substack{0 \leq i,k < n\\0 \leq j,l < p}} | \bm{D^s}_{i, k} - \bm{D^t}_{j, l}|^2 \bm{P}_{i,j} \bm{P}_{k,l}$
        }
    }\\
    &+ \rho \medspace \big(
    \tikzmarknode{unbal}{
        \highlight{red}{
            $\text{KL}(\bm{P}_{\# 1} \otimes \bm{P}_{\# 1}
        \vert \bm{w^s} \otimes \bm{w^s})
        + \text{KL}(\bm{P}_{\# 2} \otimes \bm{P}_{\# 2}
        \vert \bm{w^t} \otimes \bm{w^t})$
        }
    }
    \big)
    + \varepsilon \medspace
    \tikzmarknode{entr}{
        \highlight{gray}{
            $E(\bm{P})$
        }
    }
    \end{split}
\end{equation}
\begin{tikzpicture}[overlay,remember picture,>=stealth,nodes={align=left,inner ysep=1pt},<-]
    \path (wass.north) ++ (0.6,0.8em) node[anchor=south east,color=blue!65] (wasstext){Wasserstein loss $\text{L}_{\text{W}}(\bm{P})$};
    \draw [color=blue!85](wass.north) |- ([xshift=-0.3ex,color=blue!85]wasstext.south west);
    \path (gw.north) ++ (-2.4,2.0em) node[anchor=north west,color=PineGreen!85] (gwtext){Gromov-Wasserstein loss $\text{L}_{\text{GW}}(\bm{P})$};
    \draw [color=PineGreen](gw.north) |- ([xshift=-0.3ex,color=PineGreen]gwtext.south east);
    \path (unbal.north) ++ (-4.8,-2em) node[anchor=north west,color=red!65] (unbaltext){Marginal constraints $\text{L}_{\text{U}}(\bm{P})$};
    \draw [color=red!85](unbal.south) |- ([xshift=-0.3ex,color=red!85]unbaltext.south west);
    \path (entr.north) ++ (-2.1,-2em) node[anchor=north west,color=black!85] (entrtext){Entropy};
    \draw [color=black](entr.south) |- ([xshift=-0.3ex,color=black]entrtext.south west);
\end{tikzpicture}
\vspace*{1\baselineskip}

where $\text{L}_{\text{W}}(\bm{P})$ matches vertices with similar features,
$\text{L}_{\text{GW}}(\bm{P})$ penalizes changes in geometry
and $\text{L}_{\text{U}}(\bm{P})$ fosters matching all parts of the source and target distributions. Throughout this paper, we refer to relaxing the hard marginal constraints of the underlying OT problem into soft ones as \textit{unbalancing}.
Here, $\bm{P}_{\# 1} \triangleq (\sum_j \bm{P}_{i,j})_{0 \leq i < n}$ denotes the first marginal distribution of $\bm{P}$,
and $\bm{P}_{\# 2} \triangleq (\sum_i \bm{P}_{i,j})_{0 \leq j < p}$ the second marginal distribution of $\bm{P}$. The notation $\otimes$ represents the Kronecker product between two vectors or two matrices. $\text{KL}(\cdot|\cdot)$ denotes the Kullback Leibler divergence, which is a typical choice to measure the discrepancy between two measures in the context of unbalanced optimal transport \cite{liero_optimal_2018}.
The last term $E(\bm{P}) \triangleq \text{KL}\big(\bm{P} \otimes \bm{P} | (\bm{w^s} \otimes \bm{w^t}) \otimes (\bm{w^s} \otimes \bm{w^t})\big)$ is mainly introduced for computational purposes, as it helps accelerate the approximation scheme of the optimisation problem. Typically, it is used in combination with a small value of $\varepsilon$, so that the impact of other terms is not diluted. On the other hand, the parameters $\alpha$ and $\rho$ offer control over two other aspects of the problem: while $\alpha$ realizes a trade-off between the impact of different features and different geometries in the resulting alignment, $\rho$ controls the amount of mass transported by penalizing configurations such that the marginal distributions of the transportation plan $\bm{P}$ are far from the prior weights $\bm{w^s}$ and $\bm{w^t}$. This potentially helps adapting the size of areas where either the signal or the geometry differs too much between source and target.

Eventually, we define $\bm{\mathcal X^s} \triangleq (\bm{F^s}, \bm{D^s}, \bm{w^s})$ and
$\bm{\mathcal X^t} \triangleq (\bm{F^t}, \bm{D^t}, \bm{w^t})$,
and seek to derive an optimal coupling $\bm{P} \in \mathbb R^{n, p}$ minimizing
\begin{equation}
    \label{eq:fugw}
    \begin{split}
        \text{FUGW}(\bm{\mathcal X^s}, \bm{\mathcal X^t})
        &\triangleq \inf_{\bm{P} \geq 0} L_{\theta}(\bm{P})
    \end{split}
\end{equation}
This can be seen as a natural combination of the fused GW \cite{vayer_fused_2018} and the unbalanced GW \cite{sejourne_unbalanced_2021} distances. To the best of our knowledge, it has never been considered in the literature.
Following \cite{sejourne_unbalanced_2021}, we approximate FUGW via a lower bound. It involves solving a minimization problem with respect to two independent couplings: Using a Block-Coordinate Descent (BCD) scheme, we fix a coupling and minimize with respect to the other. This allows us to always be dealing with linear problems instead of a quadratic one. Eventually, each BCD iteration consists in alternatively solving two entropic unbalanced OT problems, whose solutions can be approximated using the scaling algorithm \cite{chizat_unbalanced_2019}. Details concerning the lower bound as well as the corresponding BCD iteration can be found in the Appendix (see Alg.~\ref{alg:lbfugw}).




\paragraph{Toy example illustrating the unbalancing property}

\begin{figure}[t]
    \centering
    \includegraphics[width=1\columnwidth]{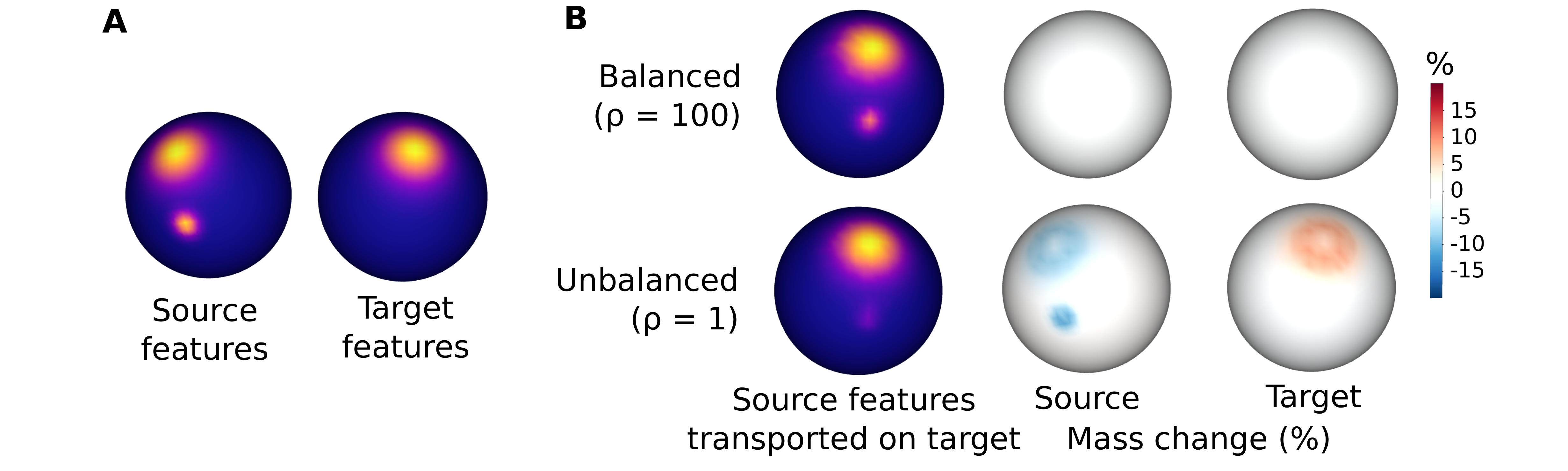}
    \caption{
        \textbf{Unbalancing helps accounting for idiosyncrasies of the source and target signals}
        When trying to align the source and target signals (Panel A), the classical balanced setup (Panel B, top row) transports all parts of the source signal even if they have no counterpart in the target signal.
        In the unbalanced setup (Panel B, bottom row), less source-only signal is transported: in particular, less mass is transported from the source's small blob onto the target (Panel B, middle column).
    }
    \label{fig:toy_example}
\end{figure}

As exemplified in Figure \ref{fig:intro}, brain responses elicited by the same stimulus vary greatly between individuals.
Figure \ref{fig:toy_example} illustrates a similar yet simplified version of this problem, where the goal is to align two different signals  supported on the same spherical meshes. 
In this example, for each of the $n=p=3200$ vertices, the feature is simply a scalar. 
On the source mesh, the signal is constituted of two von Mises density functions that differ by their concentration (large and small), while on the target mesh, only the large one is present, but at a different location. 
We use the optimal coupling matrix $\bm{P}$ obtained from Eq.~\ref{eq:fugw} to transport the source signal on the target mesh.
As shown in Figure~\ref{fig:toy_example}.B, the parameter $\rho$ allows to control the mass transferred from source to target. 
When $\rho=100$, we approach the solution of the fused GW problem. Consequently, we observe the second mode on the target when transporting the source signal.
When the mass control is weaker ($\rho=1$), the smaller blob is partly removed because it has no counterpart in the target configuration, making the transport ill-posed.
%
%

\paragraph{Barycenters}
Barycenters represent common patterns across samples. 
Their role is instrumental in identifying a unique target for aligning a given group of individuals.
%
%
As seen in Fig. \ref{fig:intro}, the vertex-wise group average does not usually provide well-contrasted maps. 
Inspired by the success of the GW distance when estimating the barycenter of structured objects ~\cite{peyre_gromov-wasserstein_2016,vayer_fused_2018}, we use FUGW to find the barycenter
$(\bm{F^B}, \bm{D^B}) \in \mathbb R^{k, c} \times \mathbb R^{k, k}$
of all subjects $s \in \mathcal S$, as well as the corresponding couplings $\bm{P^{s,B}}$ from each subject to the barycenter. More precisely, we solve
\begin{equation}
    \label{eq:barycenter}
    \bm{\mathcal X^B} = (\bm{F^B}, \bm{D^B}, \bm{w^B}) \in \argmin_{\bm{\mathcal X}} \sum_{s \in \mathcal S} \text{FUGW}(\bm{\mathcal X^s}, \bm{\mathcal X}),
\end{equation}
where we set the weights $\bm{w_B}$ to be the uniform distribution. 
By construction, the resulting barycenter benefits from the advantages of FUGW, i.e. equilibrium between geometry-preserving and feature-matching properties, while not forcing hard marginal constraints. The FUGW barycenter is estimated using a Block-Coordinate Descent (BCD) algorithm that consists in alternatively \textit{(i)} minimizing the OT plans $\bm{P^{s,B}}$ for each FUGW computation in \eqref{eq:barycenter} with fixed $\bm{\mathcal X^B}$ and \textit{(ii)} updating the barycenter $\bm{\mathcal X^B}$ through a closed form with fixed $\bm{P^{s,B}}$. See Alg. \ref{alg:fugw_barycenter} for more details. 

\section{Numerical experiments}


We design three experiments to assess the performance of FUGW.
In Experiments 1 and 2, we are interested in assessing if aligning pairs of individuals with FUGW increases correlation between subjects compared to a baseline correlation. We also compare the ensuing gains with those obtained when using the competing method MSM \cite{robinson_msm_2014, robinson_multimodal_2018} to align subjects.
In Experiment 3, we derive a barycenter of individuals
and assess its ability to capture fine-grained
details compared to classical methods.

\paragraph{Dataset}
\label{par:dataset}
In all three experiments, we leverage data from the Individual Brain Charting dataset \cite{ibc}.
It is a longitudinal study on 12 human subjects,
comprising 400 fMRI maps per subject collected
on a wide variety of stimuli (motor, visual, auditory, theory of mind,
language, mathematics, emotions, and more), movie-watching data, T1-weighted maps, as well as other features such as
retinotopy which we don't use in this work.
We leverage these 400 fMRI maps.
%
The training, validation and test sets respectively comprise
326, 43 and 30 contrast maps acquired for each individual of the dataset.
Tasks and MRI sessions differ between each of the sets.
More details, including preprocessing, are provided in Supplementary Materials.

\paragraph{Baseline alignment correlation}
For each pair of individuals $(s, t)$ under study,
and for each fMRI contrast $c$ in the test set,
we compute the Pearson correlation $\text{corr}(\bm{F^s}_{\cdot, c}, \bm{F^t}_{\cdot, c})$
after these maps have been projected onto a common surface anatomy
(in this case, \emph{fsaverage5} mesh). Throughout this work, such computations are made for each hemisphere separately.

\paragraph{Experiment 1 - Aligning pairs of humans with the same anatomy}
For each pair $(s, t)$ under study, we derive an alignment $\bm{P^{s,t}} \in \mathbb R^{n \times p}$ using FUGW on a set of training features.
In this experiment, source and target data lie on the same anatomical mesh (\emph{fsaverage5}), and $n = p = 10\ 240$ for each hemisphere. Since each hemisphere's mesh is connected, we align one hemisphere at a time.

Computed couplings are used to align contrast maps of a the validation set from the source subject onto the target subject. Indeed, one can define
$\phi_{s \rightarrow t} \colon \bm{X} \in \mathbb R^{n \times q}
\mapsto \big((\bm{P^{s,t}})^T \bm{X}\big) \oslash \bs{P^{s,t}}_{\#2} \in \mathbb R^{p \times q}$
where $\oslash$ represents the element-wise division. $\phi_{s \rightarrow t}$ transports any matrix of features from the source mesh to the target mesh.
We measure the Pearson correlation
$\text{corr}\big( \phi_{s \rightarrow t}(\bm{F^s}), \bm{F^t} \big)$
between each aligned source and target maps.

We run a similar experiment for MSM and compute the correlation gain induced on a test set by FUGW and MSM respectively.
For both models, we selected the hyper-parameters maximizing correlation gain on a validation set.
In the case of FUGW, in addition to gains in correlation, hyper-parameter selection was influenced by three other metrics that help us assess the relevance of computed couplings:

\begin{description}
    \item[Transported mass]
    For each vertex $i$ of the source subject, we compute
    $\sum\limits_{0 \leq j < p} \bm{P^{s,t}}_{i, j}$

    \item[Vertex displacement]
    Taking advantage of the fact that the source
    and target anatomies are the same, we define $\bm{D} = \bm{D^s} = \bm{D^t}$ and compute for each vertex $i$ of the source subject the quantity
    $\sum_j \bm{P^{s,t}}_{i, j} \cdot \bm{D}_{i, j} / \sum_j \bm{P^{s,t}}_{i,j}$,
    which measures the average geodesic distance on the cortical sheet between vertex $i$ and the vertices of the target it has been matched with

    \item[Vertex spread]
    Large values of $\varepsilon$ increase the entropy of derived couplings. To quantify this effect, and because we don't want the matching to be too blurry, we assess how much a vertex was \textit{spread}. Considering $\tilde{P_i} = \bm{P^{s,t}}_i / \sum_j \bm{P^{s,t}}_{i,j} \in \mathbb R^p$
    as a probability measure on target vertices, we
    estimate the anatomical variance of this measure by sampling $q$ pairs $(j_q, k_q)$ of $\tilde{P_i}$ and computing their average geodesic distance
    $\frac{1}{q} \sum\limits_{j_q, k_q} \bm{D}_{j_q, k_q}$

\end{description}

\paragraph{Experiment 2 - Aligning pairs of humans with individual anatomies}
We perform a second alignment experiment, this time using individual meshes instead of an anatomical template.
Importantly, in this case,  there is no possibility to compare FUGW with baseline methods, since those cannot handle this case.
%

However, individual meshes are significantly larger than the
common anatomical template used in Experiment 1 ($n \approx m \approx$ 160k vs. 10k previously),
resulting in couplings too large to fit on GPUs -- for reference,
a coupling of size 10k $\times$ 10k already weights ~400Mo on disk.
We thus reduce the size of the source and target data by
clustering them into 10k small connected clusters using Ward's algorithm \cite{thirion:2014}.
More details are given in supplementary section A.4.

\paragraph{Experiment 3 - Comparing FUGW barycenters with usual group analysis}
\label{par:barycenter}

Since it is very difficult to estimate the barycentric mesh, we force it to be equal to the \emph{fsaverage5} template. Empirically, this we force the distance matrix $\bm{D^B}$ to be equal to that of \emph{fsaverage5}, and only estimate the functional barycenter $\bm{F^B}$.
We initialize it with the mean of $(\bm{F^s})_{s \in S}$
and derive $\bm{F^B}$ and $(\bm{P^{s,B}})_{s \in S}$ from problem \ref{eq:barycenter}.

Then, for a given stimulus $c$, we compute its projection onto the barycenter for each subject. We use these projections to compute two maps of interest:
\textit{(i)} $\bm{M_{B,c}}$ the mean of projected contrast maps across subjects
and \textit{(ii)} $\bm{T_{B,c}}$ the t-statistic (for each vertex) of projected maps.
We compare these two maps with their unaligned counterparts $\bm{M_{0,c}}$ and $\bm{T_{0,c}}$ respectively.

\begin{figure}[ht]
    \begin{minipage}{.5\linewidth}
        \begin{equation*}
            \bm{M_{B,c}} \triangleq \frac{1}{|S|} \sum_{s \in S} \phi_{s \rightarrow t}(\bm{F^s}_{\cdot, c})
        \end{equation*}
    \end{minipage}
    \hfil
    \begin{minipage}{.5\linewidth}
        \begin{equation*}
            \bm{T_{B,c}} \triangleq \text{t-statistic} \Big( \big( \phi_{s \rightarrow t}(\bm{F^s}_{\cdot, c}) \big)_{s \in S} \Big)
        \end{equation*}
    \end{minipage}
\end{figure}
\begin{figure}[ht]
    \begin{minipage}{.5\linewidth}
        \begin{equation*}
            \bm{M_{0,c}} \triangleq \frac{1}{|S|} \sum_{s \in S} \bm{F^s}_{\cdot, c}
        \end{equation*}
    \end{minipage}
    \hfil
    \begin{minipage}{.5\linewidth}
        \begin{equation*}
            \bm{T_{0,c}} \triangleq \text{t-statistic} \Big( (\bm{F^s}_{\cdot, c})_{s \in S} \Big)
        \end{equation*}
    \end{minipage}
\end{figure}

The first map helps us to qualitatively evaluate the precision of FUGW alignments and barycenter. The second one is classically used to infer the existence of areas of the brain that respond to specific stimuli. We assess whether FUGW helps find the same clusters of vertices. Eventually, we quantify the number of vertices significantly activated or deactivated with and without alignment respectively.

\section{Results}


\subsection{Experiment 1 - Template anatomy}

\paragraph{Aligning subjects on a fixed mesh}

\begin{figure}[ht!]
    \centering
    \includegraphics[width=0.49\columnwidth]{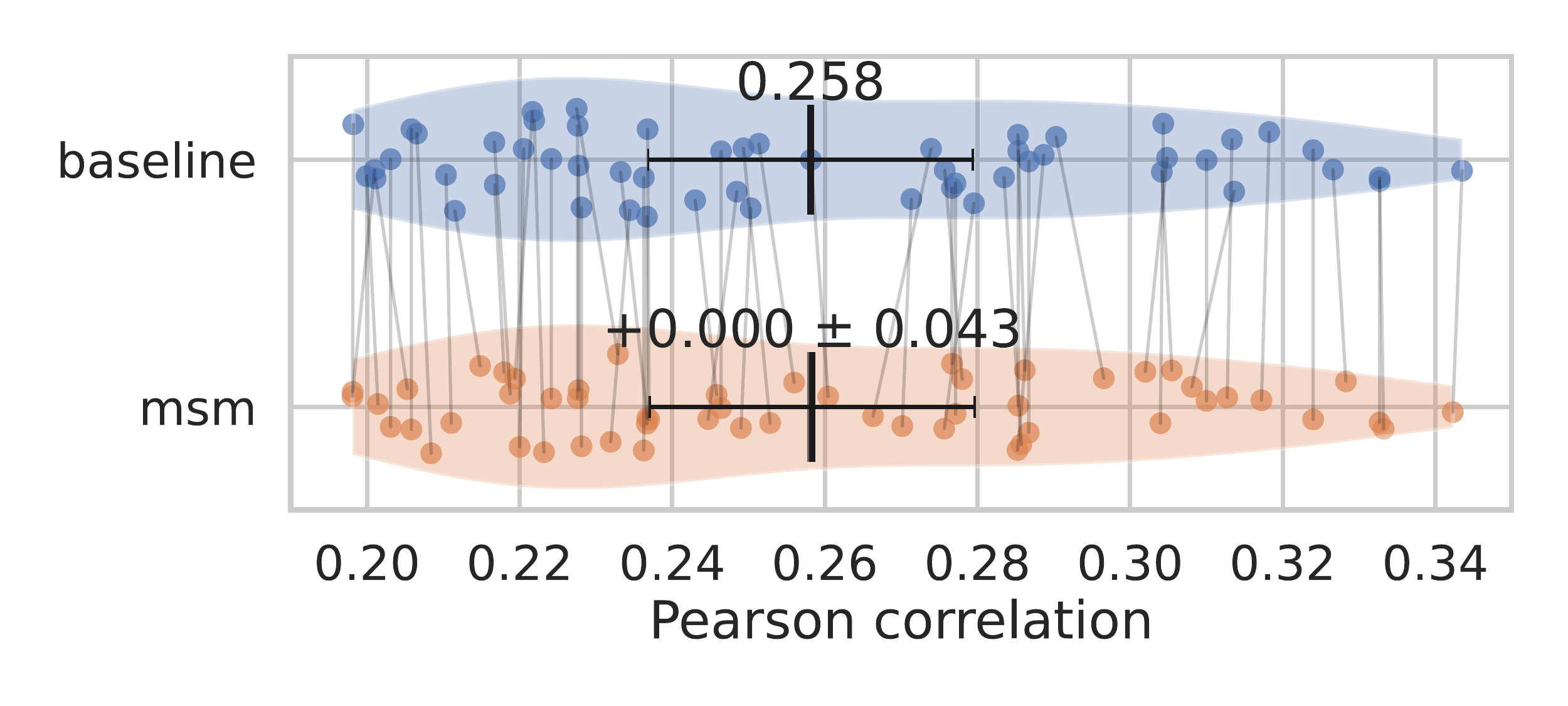}
    \includegraphics[width=0.49\columnwidth]{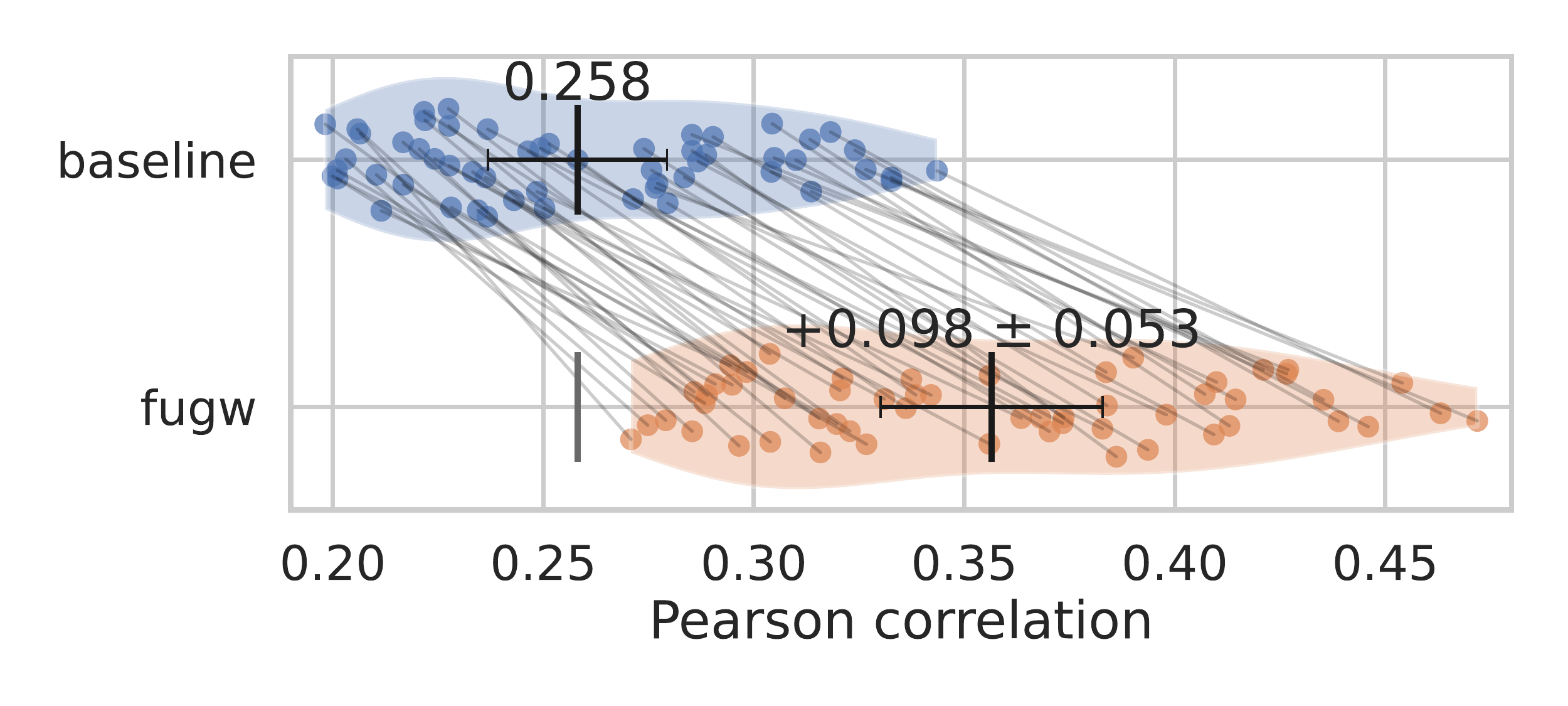}
    \caption{
        \textbf{Comparison of gains in correlation after inter-subject alignment}
        For each pair of source and target subjects
        of the dataset, we compute the average Pearson correlation between
        30 test contrasts, leading to the (baseline) correspondence score,  
        and compare it with that of the same contrast maps
        aligned with either MSM (left) or FUGW (right).
        Correlation gains are much better for FUGW.
    }
    \label{fig:gain_comparisions_fsaverage5}
\end{figure}

We set $\alpha = 0.5$, $\rho = 1$ and $\varepsilon = 10^{-3}$.
Pearson correlation between source and target contrast maps
is systematically and significantly increased when aligned using FUGW, as illustrated in Figure \ref{fig:gain_comparisions_fsaverage5} where correlation grows by almost 40\% from $0.258$ to $0.356$.

We also varied training sets by selecting subsets of training contrasts and find that similar performance
on the test set can be achieved regardless of the training data (see Supplementary section \ref{sec:control_experiments} and in particular Supplementary Table \ref{tab:varying_training_sets}). 

\paragraph{Hyper-parameters selection}
\label{par:params_selection}

\begin{figure}[ht]
    \centering
    \hspace*{-5mm}
    \includegraphics[width=1.05\columnwidth]{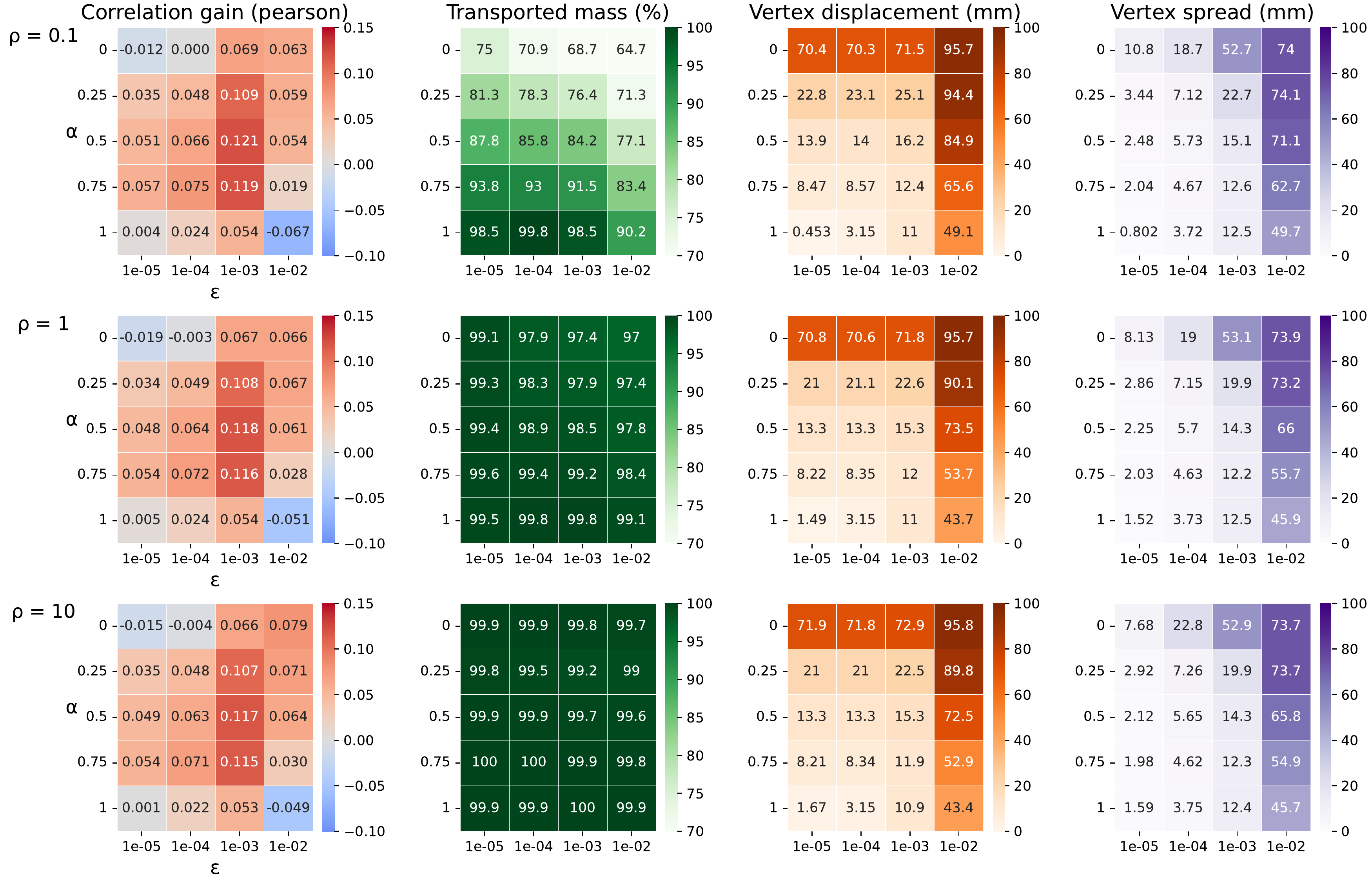}
    \caption{
        \textbf{Exploring hyper-parameter space to find relevant couplings}
        Given a transport plan aligning a source and target subject,
        we evaluate how much this coupling
        (left) improves correlation between unseen contrast maps
        of the two subjects,
        (center left) actually transports data,
        (center right) moves vertices far from their original location on the cortical surface
        and (right) spreads vertices on the cortical sheet.
        We seek plans that maximize correlation gain, while keeping spread and displacement low enough.
    }
    \label{fig:cv_metrics}
\end{figure}

Hyper-parameters used to obtain these results were chosen
after running a grid search on $\alpha$, $\varepsilon$ and $\rho$
and evaluating it on the validation dataset.
Computation took about 100 hours using 4 Tesla V100-DGXS-32GB GPUs. More precisely, it takes about 4 minutes to compute one coupling between a source and target 10k-vertex hemisphere on a single GPU, when the solver was set to run 10 BCD and 400 Sinkhorn iterations. In comparison, MSM takes about the same time on Intel(R) Xeon(R) CPU E5-2698 v4 @ 2.20GHz CPUs.
%
Results are reported in Figure \ref{fig:cv_metrics}
and provide multiple insights concerning FUGW.

Firstly, without anatomical constraint ($\alpha = 0$),
source vertices can be matched with target vertices
that are arbitrarily far on the cortical sheet.
Even though this can significantly increase correlation, it also
results in very high vertex displacement values (up to $100mm$).
Such couplings are not anatomically plausible.
Secondly, without functional information ($\alpha = 1$),
couplings recover a nearly flawless matching between source and target meshes,
so that, when $\varepsilon = 10^{-5}$
(ie when we force couplings to find single-vertex-to-single-vertex matches),
vertex displacement and spread are close to 0 and correlation is unchanged.
Fusing both constraints ($0 < \alpha < 1$)
yields the largest gains in correlation while allowing to compute
anatomically plausible reorganizations the cortical sheet between subjects.

The impact of $\rho$ (controlling marginal penalizations) on correlation seems modest, with a slight tendency of increased correlation in unbalanced problems (low $\rho$).

Finally, it is worth noting that a relatively wide range of $\alpha$ and $\rho$ yield comparable gains. 
The fact that FUGW performance is weakly sensitive to hyper-parameters makes it a good off-the-shelf tool for neuroscientists who wish to derive inter-individual alignments.
However, $\varepsilon$ is of dramatic importance in computed results and should be chosen carefully. 
Vertex spread is a useful metric to choose sensible values of $\varepsilon$; for human data one might consider that it should not exceed $20mm$.

\begin{figure}[!th]
    \centering
    \includegraphics[width=1\columnwidth]{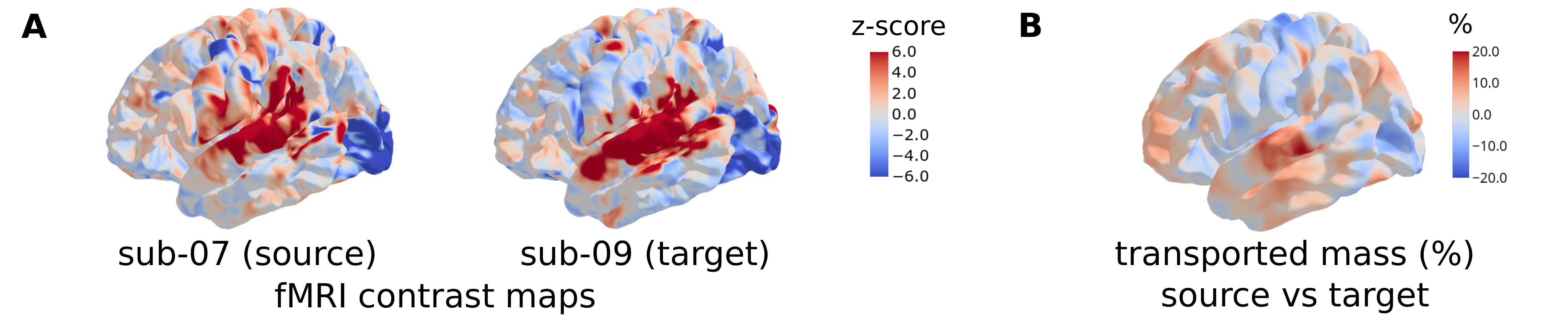}
    \caption{
        \textbf{Transported mass indicates areas which have to be resized between subjects}
        (Panel A) We show a contrast map from the test set which displays areas showing stronger activation during auditory tasks versus equivalent visual tasks. It shows much more
        anterior activations on the target subject compared to the
        source subject.
        This is consistent with the observation that more mass is present in anterior auditory areas of the source subject
        than in the target subject (Panel B).
    }
    \label{fig:transported_mass}
\end{figure}

\paragraph{Mass redistribution in unbalanced couplings}
Unbalanced couplings provide additional information about how functional areas might differ in size between pairs of individuals. 
This is illustrated in Figure \ref{fig:transported_mass},
where we observe variation in size of the auditory area
between a given pair of individuals. 
This feature is indeed captured by the difference of mass between subjects (although the displayed contrast was not part of the training set).

\subsection{Experiment 2 - Individual anatomies}

\begin{figure}[!ht]
    \centering
    \includegraphics[width=0.5\columnwidth]{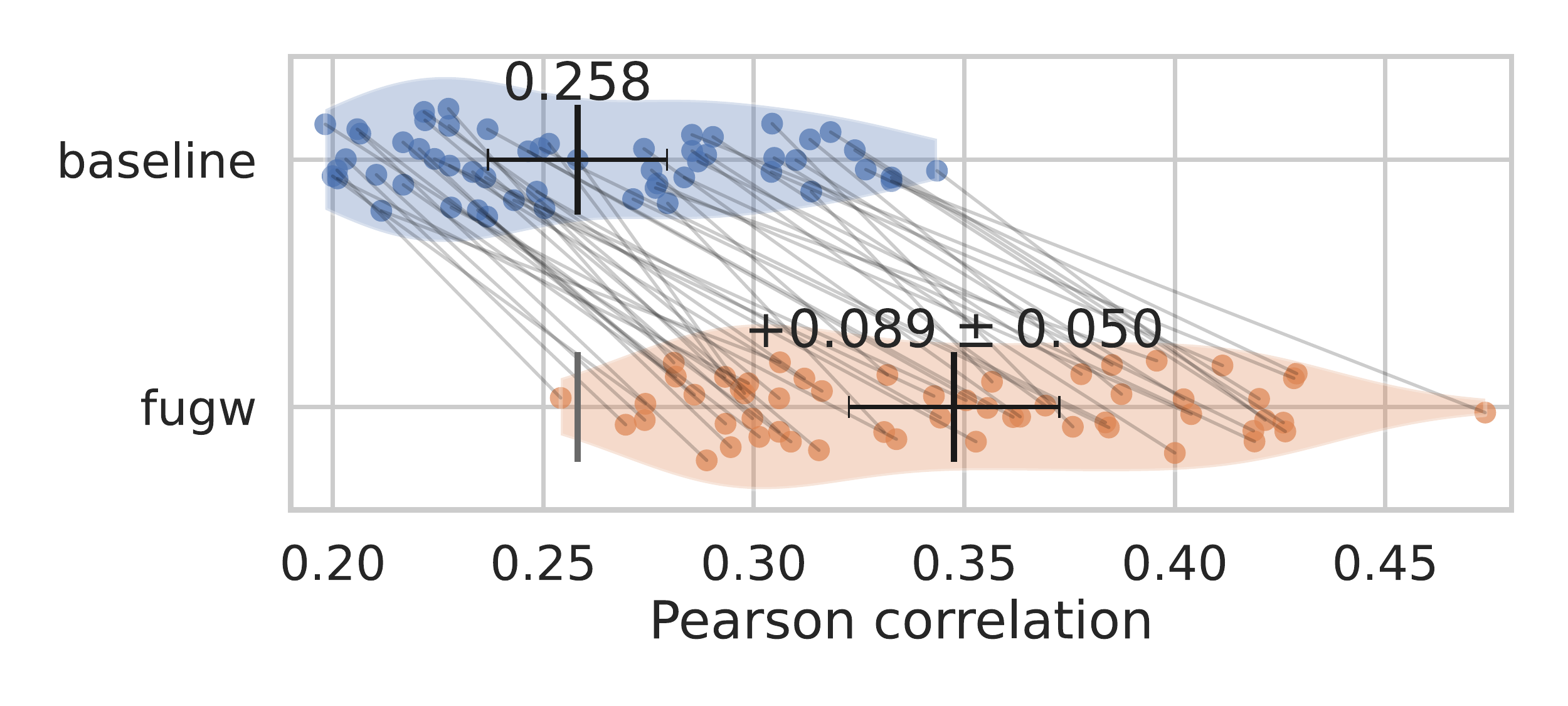}
    \caption{
        \textbf{Correlation between pairs of subjects is significantly better after alignment on individual anatomies than after projecting subjects onto a common anatomical template}
    }
    \label{fig:gain_comparisions_individual}
\end{figure}


As shown in Figure \ref{fig:gain_comparisions_individual},
we obtain correlation gains which are comparable to that of Experiment 1 (about 35\% gain) while working on individual meshes.
This tends to show that FUGW can compute meaningful alignments between pairs of individuals without the use of an anatomical template, which helps bridge most conceptual impediments listed in Section \ref{sec:introduction}.

Moreover, this opens the way for computation of simple statistics in cohorts of individuals in the absence of a template.
Indeed, one can pick an individual of the cohort and use it as a reference subject on which to transport all other individuals. We give an example in Figure \ref{fig:individual_projections}, showing that FUGW correctly preserved idiosyncrasies of each subject while transporting their functional signal in an anatomically sound way.

\subsection{Experiment 3 - Barycenter}
\begin{figure}[t]
    \centering
    \includegraphics[width=1\columnwidth]{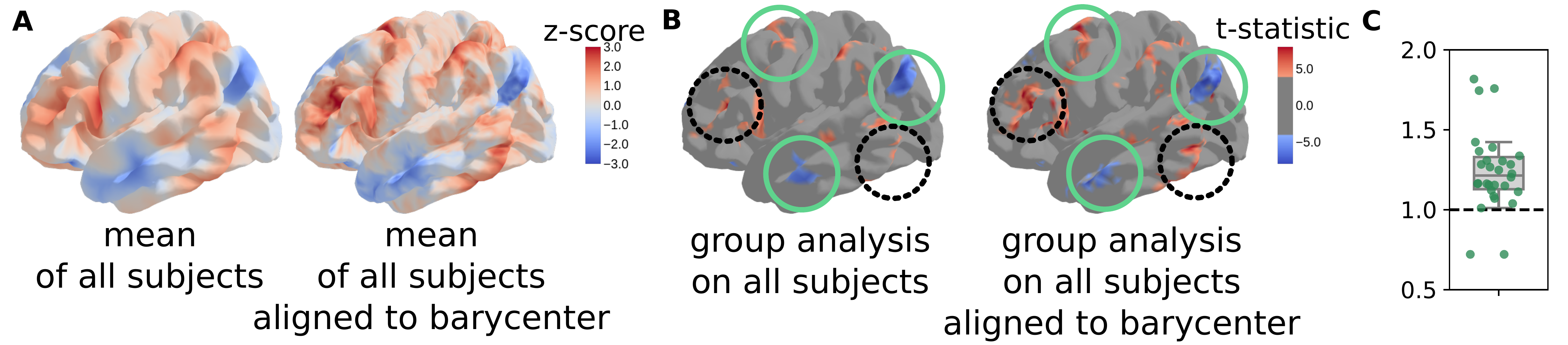}
    \caption{
        \textbf{FUGW barycenter yields much finer-grained maps than group averages}
        We study the same statistical map as in Figure \ref{fig:intro}, which contrasts
        areas of the brain involved in mathematical reasoning.
        \textbf{A}. These complex maps projected onto the barycenter and averaged show more specific activation patterns than simple group averages, especially in cortical areas exhibiting more variability, such as the prefrontal cortex.
        \textbf{B}. Deriving a t-test on aligned maps captures the same clusters as the classical approach (plain green circles), but also new clusters in areas where inter-subject variability is high (dotted black circles). Peak t-statistics are also higher with FUGW.
        \textbf{C}. Ratio of number of activated vertices ($|\text{t-statistic}| \geq 4$) with versus without alignment for each map of the test set. Our method finds significantly more of such vertices ($\text{p-value} = 3\cdot10^{-4}$).  
    }
    \label{fig:barycenter_vs_group_average}
\end{figure}

In the absence of a proper metric to quantify the correctness of a barycenter, we first qualitatively compare the functional templates obtained with and without alignment. In Figure \ref{fig:barycenter_vs_group_average}.A, we do so using brain maps taken from the test set.
%
%
We can see that the barycenter obtained with FUGW yields sharper contrasts and more fine-grained details than the barycenter obtained by per-vertex averaging.
We also display in Figure \ref{fig:barycenter_vs_group_average}.B the result of a one-sample test for the same contrast, which can readily be used for inference.
The one-sample test map obtained after alignment to the FUGW template exhibits the same supra-threshold clusters as the original approach, but also some additional spots which were likely lost due to inter-subject variability in the \emph{fsaverage5} space.
This approach is thus very useful to increase power in group inference.
We quantify this result by counting the number of supra-threshold vertices with and without alignment for each contrast map of the test set. Our alignment method significantly finds more such vertices of interest, as shown in Figure \ref{fig:barycenter_vs_group_average}.C.

\section{Discussion}

FUGW can derive meaningful couplings between pairs of subjects without the need of a pre-existing anatomical template. It is well-suited to computing barycenters of individuals, even for small cohorts.

In addition, we have shown clear evidence that FUGW yields gains that cannot be achieved by traditional diffeomorphic registration methods.
These methods impose very strong constraints to the displacement field, that may prevent reaching optimal configurations.
More deeply, this finding suggests that brain comparison ultimately requires lifting hard regularity constraints on the alignment models,  and that two human brains differ by more than a simple continuous surface deformation.
%
However, current results have not shown a strong correlation gain of unbalanced OT compared to balanced OT, likely because the cohort under study is too small. 
Leveraging datasets such as HCP \cite{hcpdata} with a larger number of subjects will help lower the standard error on correlation gain estimates.
%
In this work, we decided to rely on a predefined anatomical template (\emph{fsaverage5}) to derive functional barycenters. 
It would be interesting to investigate whether more representative anatomical templates can be learned during the process. 
This would in particular help to customize templates to different populations or species.
Additionally, using an entropic solver introduces a new hyper-parameter $\varepsilon$
that has a strong effect, but is hard to interpret.
%
Future work may replace the scaling algorithm \cite{chizat_unbalanced_2019} used here by the majorization-minimization one \cite{chapel_unbalanced_2021}, which does not require entropic smoothing. This solution can yield sparse couplings while being orders of magnitude faster, which will prove useful when computing barycenters on large cohorts.

Finally, we plan to make use of FUGW to derive alignments between human and non-human primates without anatomical priors. 
Indeed, the understanding of given brain mechanisms will benefit from more detailed invasive measurements made on other species \emph{only if} brains can be matched across species; moreover, this  raises the question of features that make the human brain unique, by identifying patterns that have no counterpart in other species.
By maximizing the functional alignment between areas, but also allowing for some regions to be massively shrunk or downright absent in one species relative to the other, the present tool could shed an objective light on the important issue of whether and how the language-related areas of the human cortical sheet map onto the architecture of non-human primate brains.

\section*{Acknowledgement}

We thank Maëliss Jallais and Thomas Moreau for their help in implementing a Python wrapper for Multimodal Surface Matching, Thibault Séjourné and Gabriel Peyré for the useful discussions. We also thank Emma Robinson and Logan Williams for helping us understand better the ropes of MSM and how to tweak it's hyper parameters.

A.T, T. Z, S. D and B.T's research has received funding from the European Union’s Horizon 2020 Framework Programme for Research and Innovation under the Specific Grant Agreement No. 945539 (Human Brain Project SGA3). It has also been supported by the the KARAIB AI chair (ANR-20-CHIA-0025-01) and the NeuroMind Inria associate team.

H. T, N. C and R. F's work is funded by the projects OTTOPIA ANR-20-CHIA-0030, 3IA Côte d’Azur Investments ANR19-P3IA-0002 of the French National Research Agency (ANR), the 3rd Programme d’Investissements d’Avenir ANR-18-EUR-0006-02, and the Chair "Business Analytics for Future Banking" sponsored by NATIXIS. This research is produced within the framework of Energy4Climate Interdisciplinary Center (E4C) of IP Paris and Ecole des Ponts ParisTech.

\newpage
\printbibliography



\newpage
\appendix

\beginsupplement

\section{Appendix}

In the following sections, we provide some additional material to ease the understanding of the underlying alignment problem as well as computational details of solutions of FUGW and FUGW barycenters.

We also show some control experiments during which we used different training data to compute pair-wise alignments
and evaluated the proportion of correlation gains that comes from mere signal smoothing.

Eventually, we give details about the IBC dataset (acquisition, preprocessing, fMRI protocols and data splitting).

\subsection{Illustration of the alignment problem}
\begin{figure}[ht]
    \centering
    \includegraphics[width=1\linewidth]{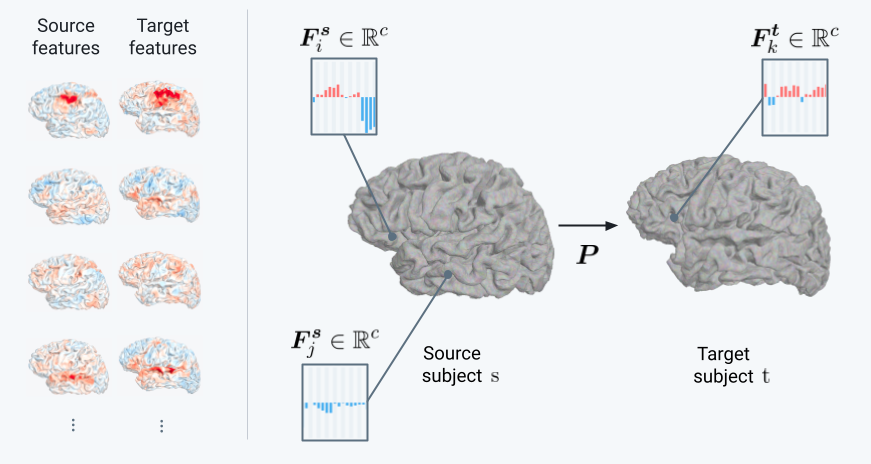}
    \caption{
        \textbf{Alignment of two brains using functional signatures}
        Using multiple maps of comparable features (left column)
        for the source and target subjects,
        we seek to derive an alignment (also referred to as a \emph{coupling})
        $\bm{P}$ that matches
        parts of the brain with similar features
        while preserving the global geometry of the cortex.
    }
    \label{fig:feature_based_alignment}
\end{figure}

We provide in Fig. \ref{fig:feature_based_alignment} a conceptual illustration of the alignment framework for a pair of subjects.

\subsection{Detailed description of FUGW estimation algorithm}

Estimating the unbalanced Gromov Wasserstein (UGW) loss is numerically sensitive to initialization, due to the non-convexity of the problem \cite{sejourne_unbalanced_2021}.
Therefore FUGW is \textit{a priori} non-convex as well, and comparably difficult to estimate.
Consequently, following \cite{sejourne_unbalanced_2021}, we instead compute a lower bound which we formulate as a bi-convex problem that relies on the joint estimation of two couplings.
%
%
We provide next a detailed derivation of this estimation procedure,
using notations introduced in section 2:
\begin{equation}
    \label{eq:lbfugw}
    \begin{split}
        \quad \text{FUGW}(\bm{\mathcal X^s}, \bm{\mathcal X^t}) = \quad
        \inf_{\substack{\bm{P}, \bm{Q} \geq 0 \\ \bm{P} = \bm{Q}}}
        \text{L}_{\theta}(\bm{P}, \bm{Q})
        \quad \geq
        \inf_{\substack{\bm{P}, \bm{Q} \geq 0 \\ m(\bm{P}) = m(\bm{Q})}}
        L_{\theta}(\bm{P}, \bm{Q})
        \quad \triangleq \quad
        \text{LB-FUGW} (\bm{\mathcal X^s}, \bm{\mathcal X^t}).
    \end{split}
\end{equation}
where $m(\bm{P}) = \sum_{i,j} \bm{P}_{i,j}$ denotes the mass of $\bm{P}$.

We give the explicit formulation of $\text{L}_{\theta}(\bm{P}, \bm{Q})$
\begin{equation}
    \label{eq:fugw_loss_two_couplings}
    \begin{split}
        \text{L}_{\theta}(\bm{P}, \bm{Q}) \triangleq
        (1 - \alpha) \enspace \text{L}_{\text{W}}(\bm{P}, \bm{Q})
        + \alpha \enspace \text{L}_{\text{GW}}(\bm{P}, \bm{Q})
        + \rho \enspace \text{L}_{\text{U}}(\bm{P}, \bm{Q})
        + \varepsilon \enspace E(\bm{P}, \bm{Q}),
    \end{split}
\end{equation}
where
\begin{itemize}
    \item $\bm{C} \triangleq \Big( ||\bm{F^s}_{i} - \bm{F^t}_j||^2_2\Big)_{i,j} \in \mathbb R^2_+$ \hfill (feature cost matrix)

    \item $\bm{G} \triangleq \Big( |\bm{D^s}_{i,j} - \bm{D^t}_{k,l}| \Big)_{i,j,k,l} \in \mathbb R^4_+$ \hfill (geometry cost matrix)

    \item $\text{L}_{\text{W}}(\bm{P}, \bm{Q}) \triangleq \langle \bm{C}, \frac{\bm{P} + \bm{Q}}{2} \rangle = \frac{1}{2} ( \sum_{i,j} \bm{C}_{i,j}\bm{P}_{i,j} + \sum_{i,j} \bm{C}_{i,j}\bm{Q}_{i,j})$ \hfill (Wasserstein)

    \item $\text{L}_{\text{GW}}(\bm{P}, \bm{Q}) \triangleq \langle \bm{G} , \bm{P} \otimes \bm{Q} \rangle = \sum_{i,j,k,l}\bm{G}_{i,j,k,l}\bm{P}_{i,j}\bm{Q}_{k,l}$ \hfill (Gromov-Wasserstein)

    \item $\text{L}_{\text{U}}(\bm{P}, \bm{Q}) \triangleq \enspace \text{KL} \Big(\bm{P}_{\# 1} \otimes \bm{Q}_{\# 1} \vert \bm{w}^s \otimes \bm{w}^s \Big) + \enspace \text{KL} \Big(\bm{P}_{\# 2} \otimes \bm{Q}_{\# 2} \vert \bm{w}^t \otimes \bm{w}^t \Big)$ \hfill (unbalancing)

    \item $E(\bm{P}, \bm{Q}) \triangleq \text{KL} \Big(\bm{P} \otimes \bm{Q} | (\bm{w^s} \otimes \bm{w^t}) \otimes (\bm{w^s} \otimes \bm{w^t}) \Big)$ \hfill (entropy)
\end{itemize}
In particular, we have $\text{L}_{\theta}(\bm{P}, \bm{P}) = \text{L}_{\theta}(\bm{P})$,
which is the objective function of FUGW introduced
in Equation 1.

\paragraph{Solver}

We provide a Python GPU-based solver for LB-FUGW,
using an approach similar to that of \cite{sejourne_unbalanced_2021},
which we recall in algorithm \ref{alg:lbfugw}. More precisely, we alternatively optimize one coupling while keeping the other fixed. This results in two entropic unbalanced OT problems in each iteration, which can be solved using the scaling algorithm \cite{chizat_unbalanced_2019}.

\begin{algorithm}[th]
    \caption{Approximation scheme for LB-FUGW}
    \label{alg:lbfugw}
    \begin{algorithmic}[1]
        \Require{$\bm{\mathcal X^s}, \bm{\mathcal X^t}, \rho, \alpha, \varepsilon$.}
        \Ensure{Pair of optimal couplings $(\bm{P}, \bm{Q})$.}
        \State Initialize: $\bm{P} = \bm{Q} = \bm{w^s} \otimes \bm{w^t} / \sqrt{m(\bm{w^s}) m(\bm{w^t})}$.
        \While{$(\bm{P}, \bm{Q})$ has not converged}
            \State Calculate: $c_P = \text{Cost}(\bm{P},  \bm{G}, \bm{C}, \bm{w^s}, \bm{w^t}, \rho, \alpha, \varepsilon)$.
            \State Update: $\bm{Q} \gets \text{Scaling}(c_P, \bm{w^s}, \bm{w^t}, \rho m(\bm{P}), \varepsilon m(\bm{P}))$.
            \State Rescale: $\bm{Q} \gets \sqrt{\frac{m(\bm{P})}{m(\bm{Q})}} \bm{Q}$.
            \State Calculate: $c_Q = \text{Cost}(\bm{Q},  \bm{G}, \bm{C}, \bm{w^s}, \bm{w^t}, \rho, \alpha, \varepsilon)$.
            \State Update: $\bm{P} \gets \text{Scaling}(c_Q, \bm{w^s}, \bm{w^t}, \rho m(\bm{Q}), \varepsilon m(\bm{Q}))$.
            \State Rescale: $\bm{P} \gets \sqrt{\frac{m(\bm{Q})}{m(\bm{P})}} \bm{P}$.
        \EndWhile
    \end{algorithmic}
\end{algorithm}

\begin{algorithm}[th]
    \caption{Scaling algorithm~\cite{chizat_unbalanced_2019}}
    \label{alg:sinkhorn}
    \begin{algorithmic}[1]
        \Require{$\bm{C}, \bm{w^s}, \bm{w^t}, \rho, \varepsilon$.}
        \Ensure{Optimal coupling $\bm{P}$.}
        \State Initialize dual vectors: $f = 0_n \in \mathbb R^n, g = 0_p \in \mathbb R^p$.
        \While{$(f,g)$ has not converged}
            \State Update: $f = -\frac{\rho}{\rho + \varepsilon} \log \sum_j \exp \big( g_j + \log \bm{w^t}_j - \frac{\bm{C}_{\cdot,j}}{\varepsilon} \big)$.
	        \State Update: $g = -\frac{\rho}{\rho + \varepsilon} \log \sum_i \exp \big( f_i + \log \bm{w^s}_i - \frac{\bm{C}_{i,\cdot}}{\varepsilon} \big)$.
        \EndWhile
        \State Calculate: $\bm{P} = (\bm{w^s} \otimes \bm{w^t}) \exp \big(f \oplus g - \frac{\bm{C}}{\varepsilon} \big)$.
    \end{algorithmic}
\end{algorithm}

\begin{algorithm}[th]
    \caption{Cost}
    \label{alg:local_cost}
    \begin{algorithmic}[1]
        \Require{$\bm{P}, \bm{G}, \bm{C}, \bm{w^s}, \bm{w^t}, \rho, \alpha, \varepsilon$.}
        \Ensure{Local cost $c$.}
        \State Calculate: $\bm{G} \otimes \bm{P} := \left( \sum_{i,j} \bm{G}_{i,j,k,l} \bm{P}_{i,j} \right)_{k,l}$.
        \State Calculate:
        \begin{equation*}
        \begin{split}
            c := \alpha \; \bm{G} \otimes \bm{P} + \frac{1 - \alpha}{2} \; \bm{C} + \rho \; \langle \log \frac{\bm{P}_{\#1}}{\bm{w^s}}, \bm{P}_{\#1} \rangle + \rho \; \langle \log \frac{\bm{P}_{\#2}}{\bm{w^t}}, \bm{P}_{\#2} \rangle + \varepsilon \; \langle \log \frac{\bm{P}}{\bm{w^s} \otimes \bm{w^t}}, \bm{P} \rangle
        \end{split}
    \end{equation*}
    \end{algorithmic}
\end{algorithm}
Here, the notations $\otimes$ and $\oplus$ denote the Kronecker product and sum, respectively. The exponential, division and logarithm operations are all element-wise. The scalar product is denoted by $\langle \cdot, \cdot \rangle$.

In practice, we observe that the two couplings of LB-FUGW are numerically equal, so it is enough to choose, for example, the first one, as alignment between source and target signals.

\subsection{Detailed description of FUGW barycenter estimation}
FUGW-barycenter algorithm, described in  Algorithm \ref{alg:fugw_barycenter}, alternates between computing mappings from subjects
to the barycenter, and updating the barycenter.
This corresponds to a block coordinate descent on the barycenter estimation.
The first step simply uses the previously introduced solver.
The second one takes advantage of the fact that the objective function introduced in \ref{eq:lbfugw} is differentiable in $\bm{F^B}$ and $\bm{D^B}$, and the two couplings of LB-FUGW are numerically equal. This yields a closed form for $\bm{F^B}$ and $\bm{D^B}$, as a function of $\bm{P^{s,B}}$ and $\bm{\mathcal X^s}$. We note that, during the barycenter estimation, the weight $\bm{w^B}$ is always fixed as uniform distribution.

\begin{algorithm}[th]
    \caption{LB-FUGW barycenter}
    \label{alg:fugw_barycenter}
    \begin{algorithmic}[1]
        \Require{$(\bm{\mathcal X^s})_{s \in \mathcal S}, \rho, \alpha, \varepsilon$.}
        \Ensure{
            Individual couplings $(\bm{P^{s, B}})_{s \in \mathcal S}$,
            barycenter $\bm{\mathcal X^B}$.
        }

        \State Initialize: $\bm{F^B} = \bm{\mathbb I_k}$; $\bm{D^B} = \bm{\mathbb 0_k}$.
        \While{$\bm{\mathcal X^B} = (\bm{F^B}, \bm{D^B}, \bm{w^B})$ has not converged}
            \State Draw $\widetilde{S}$ subset of $S$.
            \For{$s \in \widetilde{S}$}
                \Comment{fixed $\bm{\mathcal X^B}$}
                \State Align: $\bm{P^{s, B}} \gets
                \text{LB-FUGW}(\bm{\mathcal X^s}, \bm{\mathcal X^B}, \rho, \alpha, \varepsilon)$.
            \EndFor
            \State Update $\bm{F^B}$, $\bm{D^B}$:
            \Comment{fixed $\bm{P^{s, B}}$}
            \begin{equation*}
                \bm{F^B} = \frac{1}{| \widetilde{S} |} \sum_{s \in \widetilde{S}} \text{diag} \left( \frac{1}{\bm{P^{s, B}}_{\# 2}} \right) (\bm{P^{s, B}})^\top \bm{F^s} \; \text{ and } \; \bm{D_B} = \frac{1}{| \widetilde{S} |} \sum_{s \in \widetilde{S}} \frac{(\bm{P^{s, B}})^\top \bm{D^s} \bm{P^{s, B}}}{\bm{P^{s, B}}_{\# 2} (\bm{P^{s, B}}_{\# 2})^\top}.
            \end{equation*}
        \EndWhile

    \end{algorithmic}
\end{algorithm}

\subsection{Implementation details}
\label{sec:implementation_details}

\paragraph{MSM configuration}
We use the default configuration of MSM
\footnote{MSM default configuration \url{https://github.com/ecr05/MSM_HOCR/blob/master/config/basic_configs/config_standard_MSMpair}}
and vary parameter \emph{lambda} so as to obtain the best gains in correlation on the test set.
We use the same value of \emph{lambda} at each step of MSM
and eventually set it to 0.1 after a cross validated grid search.

\begin{figure}[H]
    \centering
    \begin{tabular}{cc}
    \hspace*{-5mm}
    \begin{subfigure}{.49\textwidth}
        \includegraphics[width=1\linewidth]{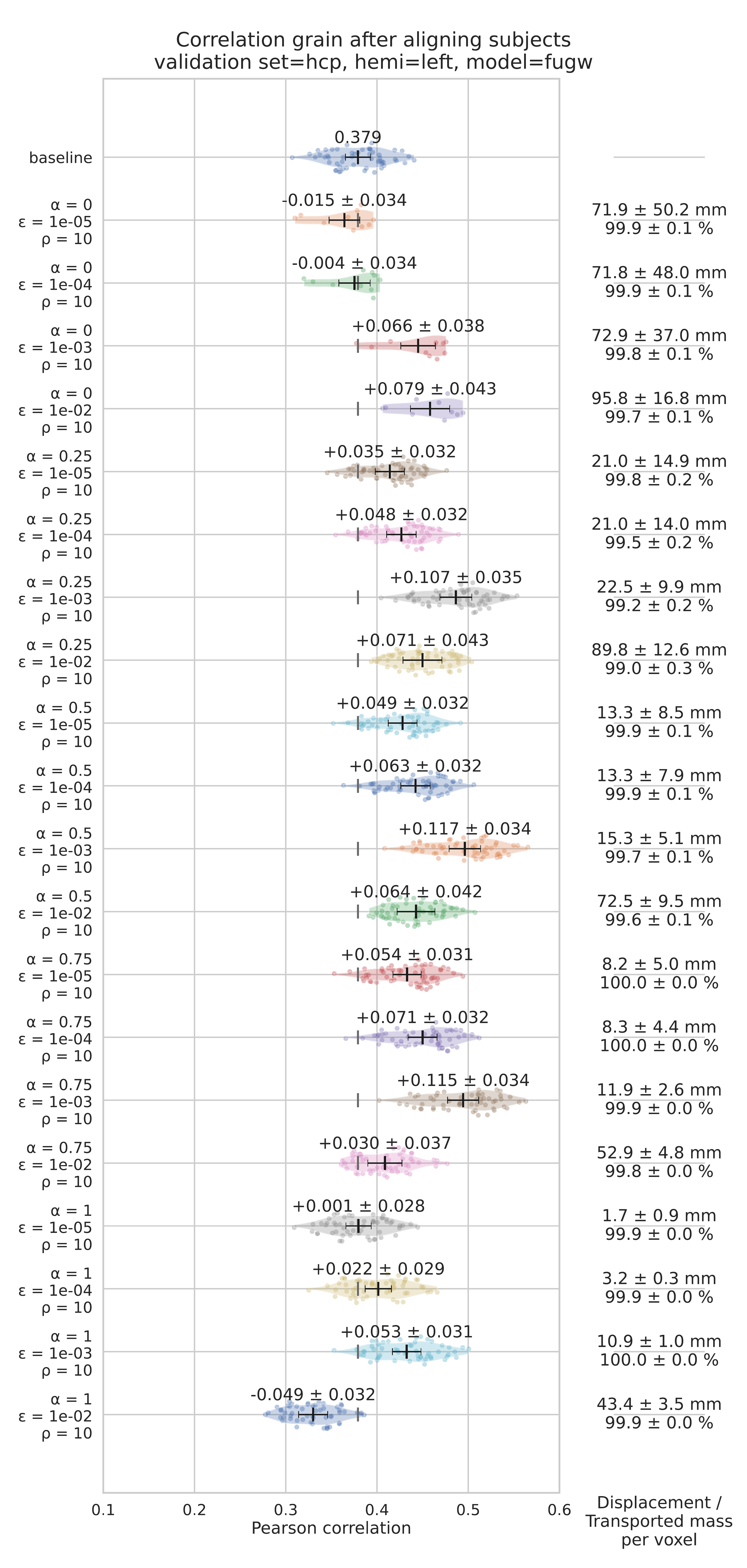}
    \end{subfigure}
    & 
    \begin{subfigure}{.49\textwidth}
        \centering
        \includegraphics[width=1\linewidth]{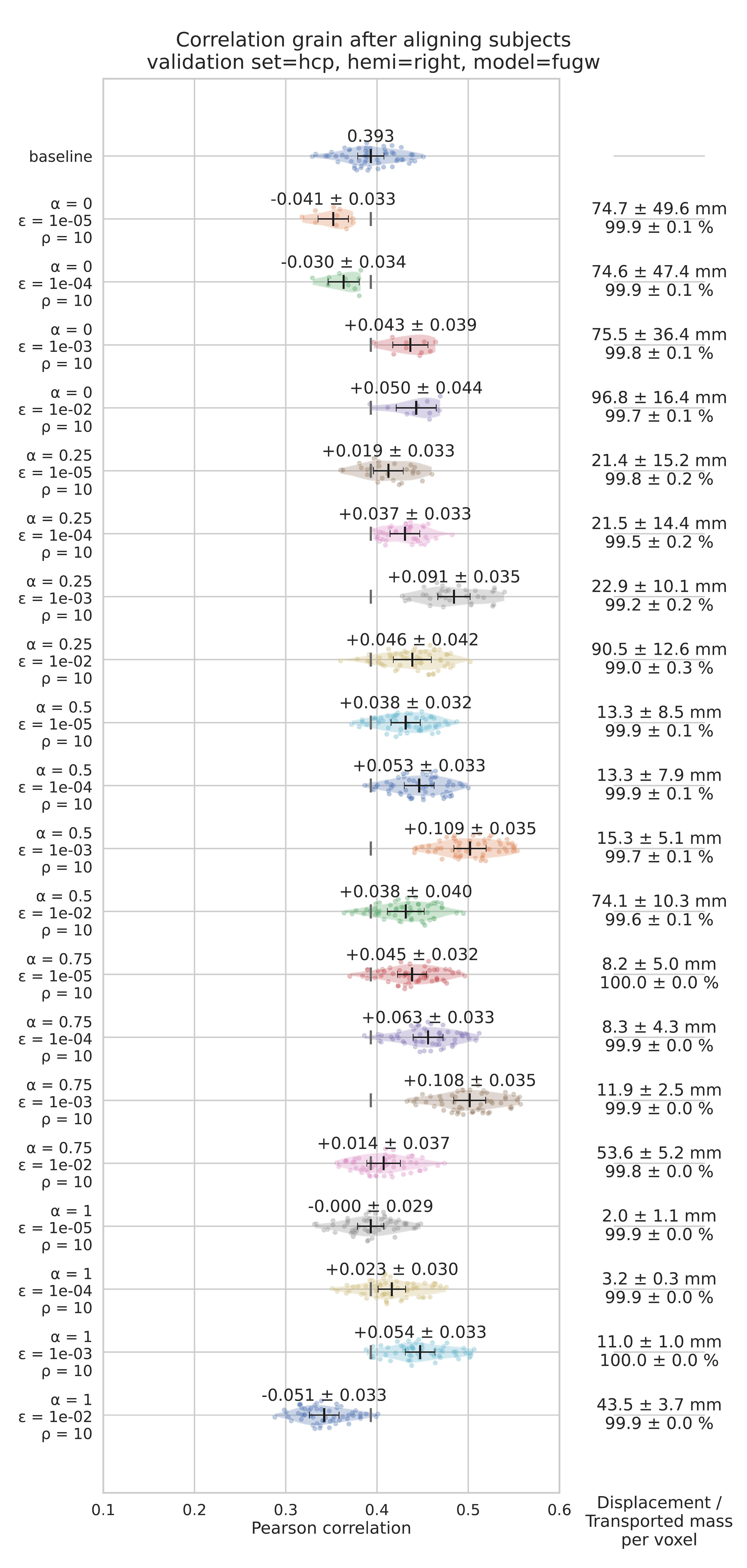}
    \end{subfigure}
    \end{tabular}
    \caption{
        \textbf{Detailed correlation gains on the validation set (HCP tasks), in the balanced case}
        Each line represents a FUGW model trained with different hyper-parameters.
        Each dot represents the mean correlation between contrast maps of the HCP protocol
        for a given pair of IBC subjects.
        We compare the average correlation with that of the baseline (top row)
        where subjects were simply projected on \emph{fsaverage5}.
        Models for the left hemisphere and right hemisphere are shown respectively on the left and right side.
    }
    \label{fig:cv_hcp_fugw}
\end{figure}

\begin{figure}[H]
    \begin{tabular}{cc}
    \hspace*{-5mm}
    \begin{subfigure}{.49\textwidth}
        \centering
        \includegraphics[width=1\linewidth]{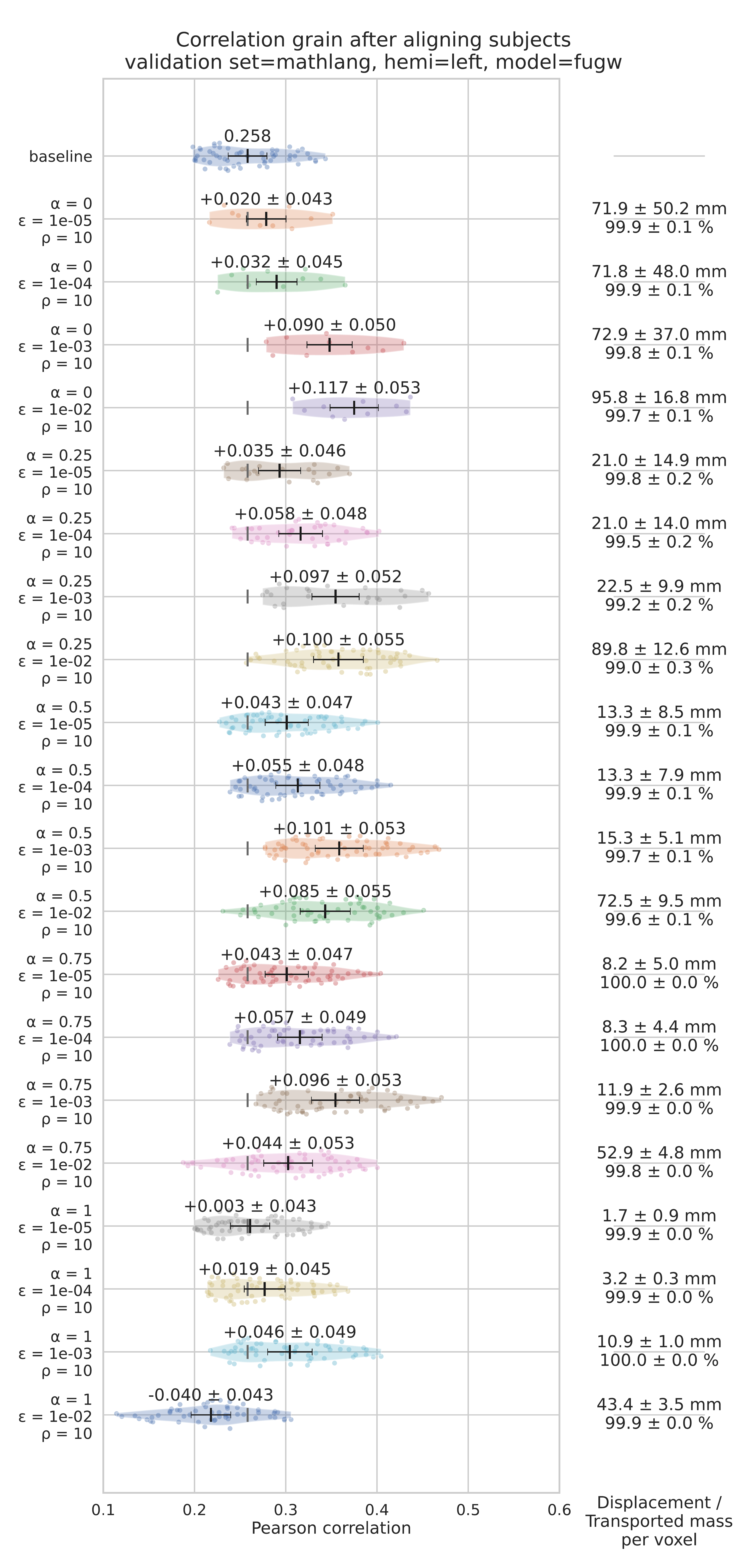}
    \end{subfigure}
    & 
    \begin{subfigure}{.49\textwidth}
        \centering
        \includegraphics[width=1\linewidth]{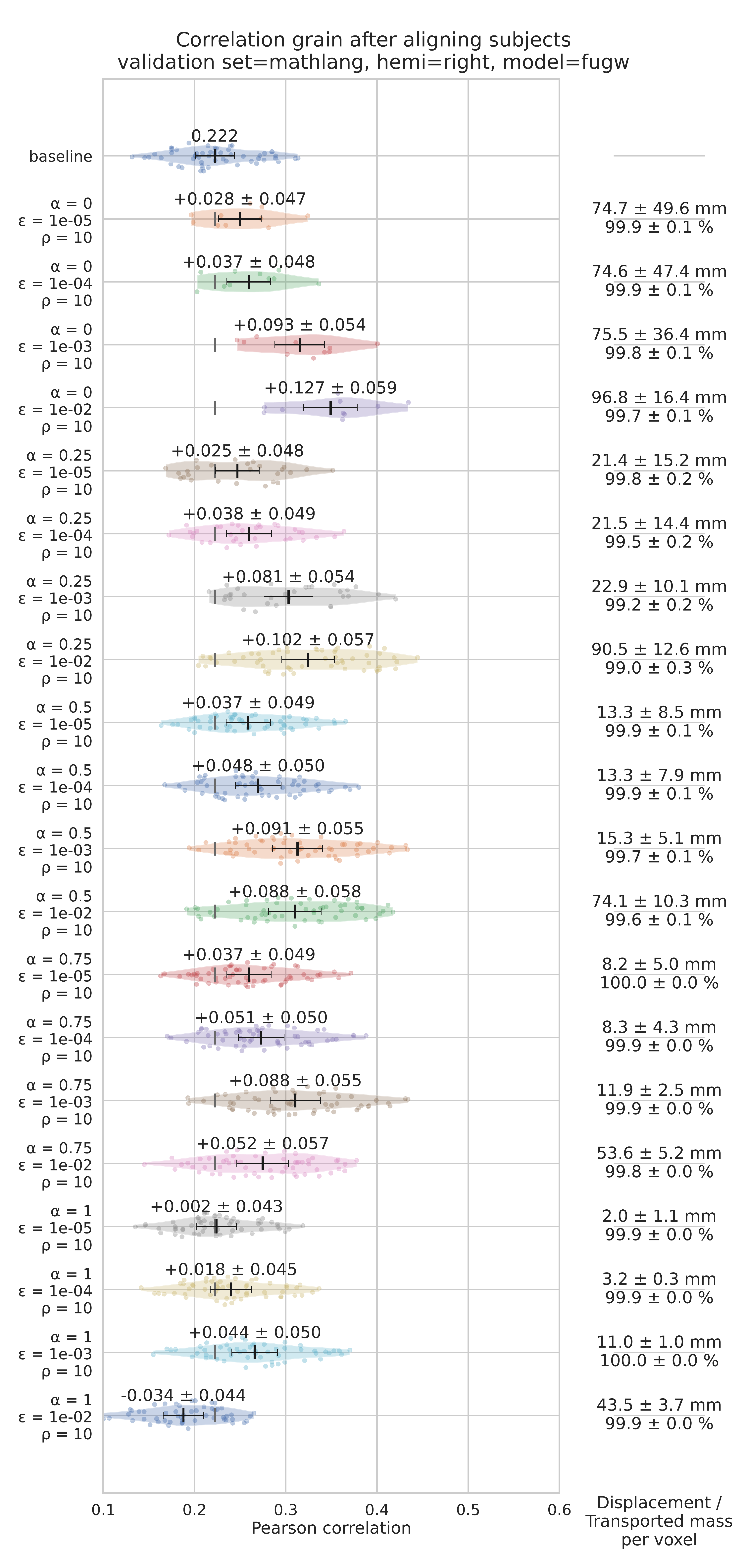}
    \end{subfigure}
    \end{tabular}
    \caption{
        \textbf{Detailed correlation gains on the test set (Mathlang tasks) in the balanced case}
        Similarly to Figure \ref{fig:cv_hcp_fugw},
        each line represents a FUGW model trained with different hyper-parameters.
        Each dot represents the mean correlation between contrast maps of the Mathlang protocol
        for a given pair of IBC subjects.
        We compare the average correlation with that of the baseline (top row)
        where subjects were simply projected on \emph{fsaverage5}.
        Models for the left hemisphere and right hemisphere are shown respectively on the left and right side.
    }
    \label{fig:cv_mathlang_fugw}
\end{figure}

\paragraph{Correlation gain variability when aligning pairs of subjects}
Figures \ref{fig:cv_hcp_fugw} and \ref{fig:cv_mathlang_fugw} show correlation gains on the validation and test sets respectively when aligning pairs of subjects from the IBC dataset. Subjects' data was previously projected onto \emph{fsaverage5}.
These figures provide us with a better understanding of the standard error and consistency of these gains. Moreover, they show that selection of the best set of hyper-parameters is robust to changing the validation data.

\newpage

\paragraph{Mesh resolution reduction}
As mentioned in the core of this paper, aligning meshes with high resolutions can lead to dealing with matrices which won't fit on GPUs. This is typically the case when trying to align two \textit{fsaverage7} hemispheres (160k vertices each) instead of \textit{fsaverage5} hemispheres (10k vertices each).

In order to reduce the number $n$ of aligned vertices, we first group them into small clusters using Ward's algorithm using a method described in \cite{thirion:2014}. In essence, this method iteratively groups adjacent vertices of a given individual based on feature similarity until $I$ clusters have been formed.
Then, for a given cluster $u_i$ of the source subject $s$, we define its functional signal $\bm{\hat{F}^s}_{u_i}$ as the mean functional signal of vertices which belong to this cluster.
Moreover, for two given clusters $u_i$ and $u_j$ of subject $s$, we define the anatomical distance $\bm{\hat{D}^s}_{u_i, u_j}$ between $u_i$ and $u_j$ as the mean geodesic distance between all pairs of vertices between the two clusters (akin to an Energy distance).
Eventually, we derive analogous objects $\bm{\hat{F}^t}$ and $\bm{\hat{D}^t}$ for the target subject $t$, and end up in a configuration comparable to that of Experiment 1.

\vspace*{-0.5\baselineskip}
\begin{figure}[ht]
    \begin{minipage}{0.5\columnwidth}
        \begin{equation*}
            \bm{\hat{F}^s}_{u_i} \triangleq
            \frac{1}{|u_i|} \sum_{k \in u_i} \bs{F^s}_{k} \in \mathbb R^c
        \end{equation*}
    \end{minipage}
    \hfil
    \begin{minipage}{0.5\columnwidth}
        \begin{equation*}
            \bm{\hat{D}^s}_{u_i, u_j} \triangleq
            \frac{1}{|u_i| \cdot |u_j|}
            \sum_{k \in u_i, l \in u_j} \bm{D}^s_{k, l}
        \end{equation*}
    \end{minipage}
\end{figure}
\vspace*{-0.5\baselineskip}

\begin{figure}[H]
    \centering
    \includegraphics[width=1\columnwidth]{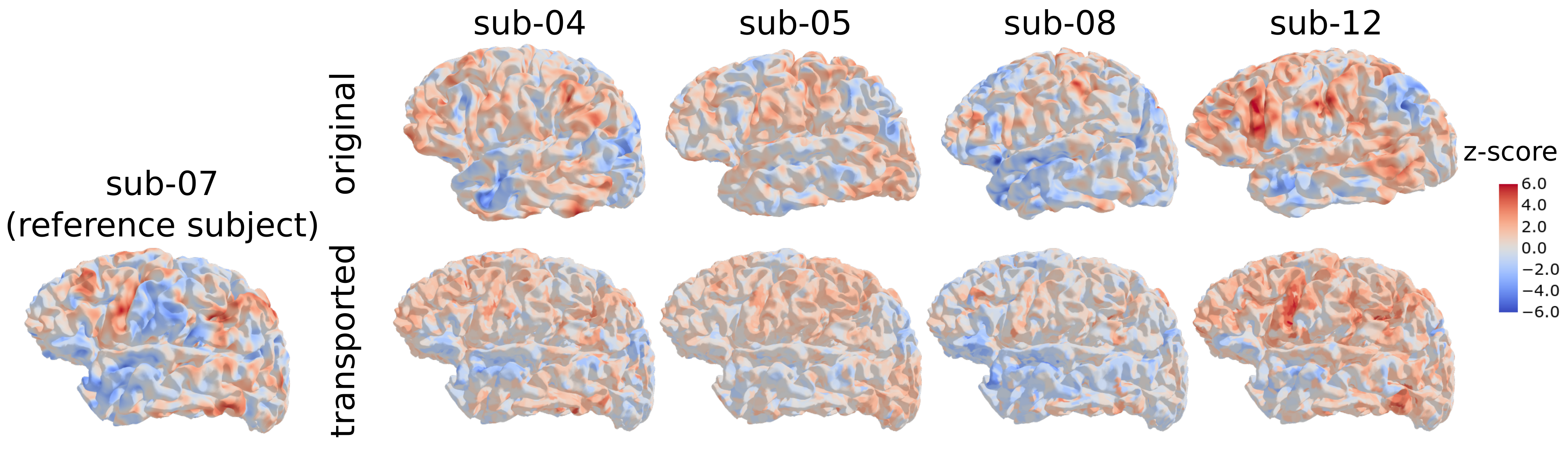}
    \caption{
        \textbf{Transporting individual maps onto a reference subject}
        FUGW can help bridge the absence of template anatomies
        and derive pairs of alignments such that all individuals
        of the cohort are comparable. We display a map taken from the test set contrasting areas activated during mathematical reasoning against areas activated for other stimuli of the protocol.
    }
    \label{fig:individual_projections}
\end{figure}

\paragraph{Alignment to individual anatomy}
We qualitatively control that alignments derived between individuals on their individual anatomies make sense in Figure \ref{fig:individual_projections}.

\subsection{Control experiments}
\label{sec:control_experiments}

\paragraph{Controlling for smoothing effect increasing correlation}
Alignments computed with FUGW are not always vertex-to-vertex alignments.
Indeed, a single vertex from the source subject $s$ can be associated with many vertices in the target subject $t$.
In fact, $\bm{P^{s,t}}_i$ represents the relative importance of each match.
The hyper-parameter $\varepsilon$ controls the entropy of $\bm{P^{s,t}}$,
which is in direct link with the spread of vertices
that we use as a measure for how many target vertices are matched with source vertices.

\begin{figure}[t]
    \centering
    \includegraphics[width=0.49\columnwidth]{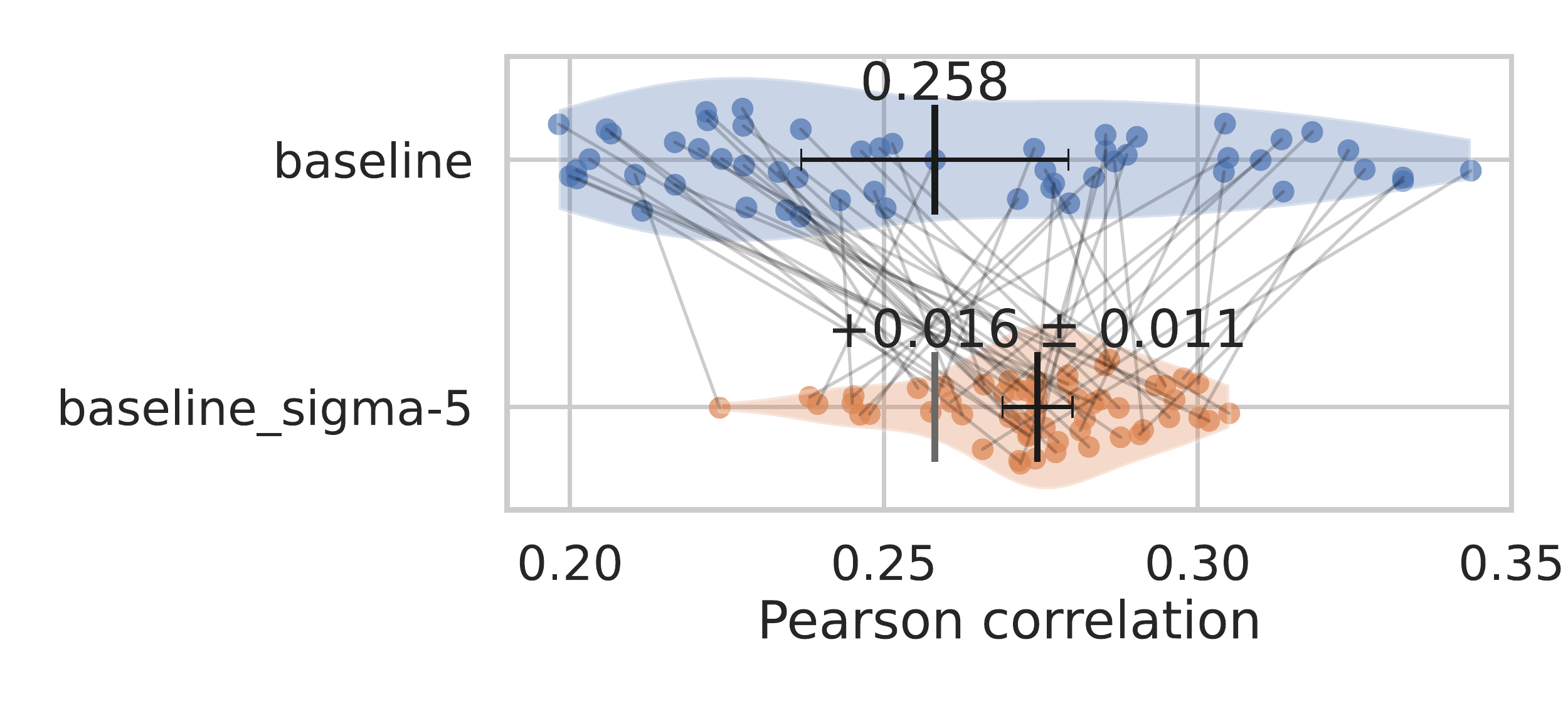}
    \includegraphics[width=0.49\columnwidth]{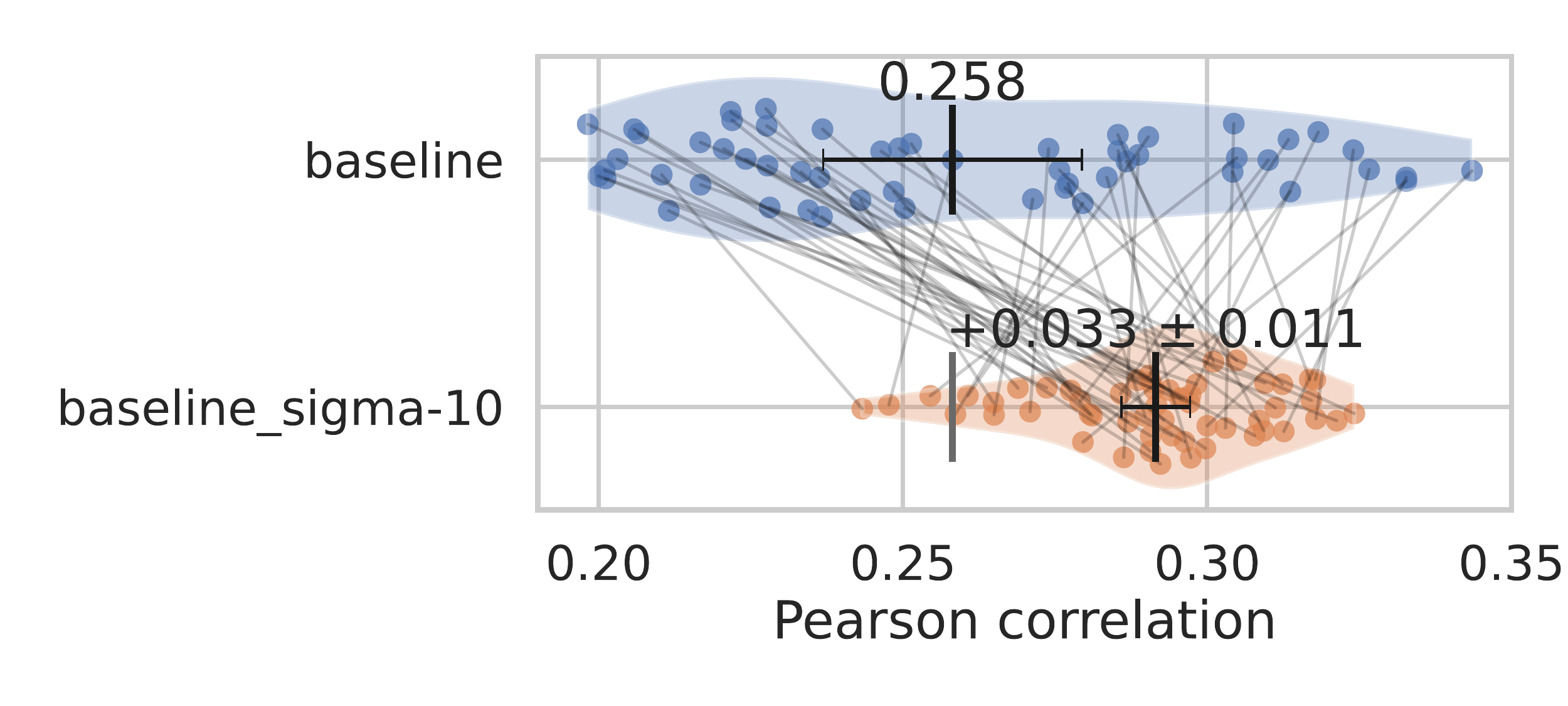}
    \includegraphics[width=0.49\columnwidth]{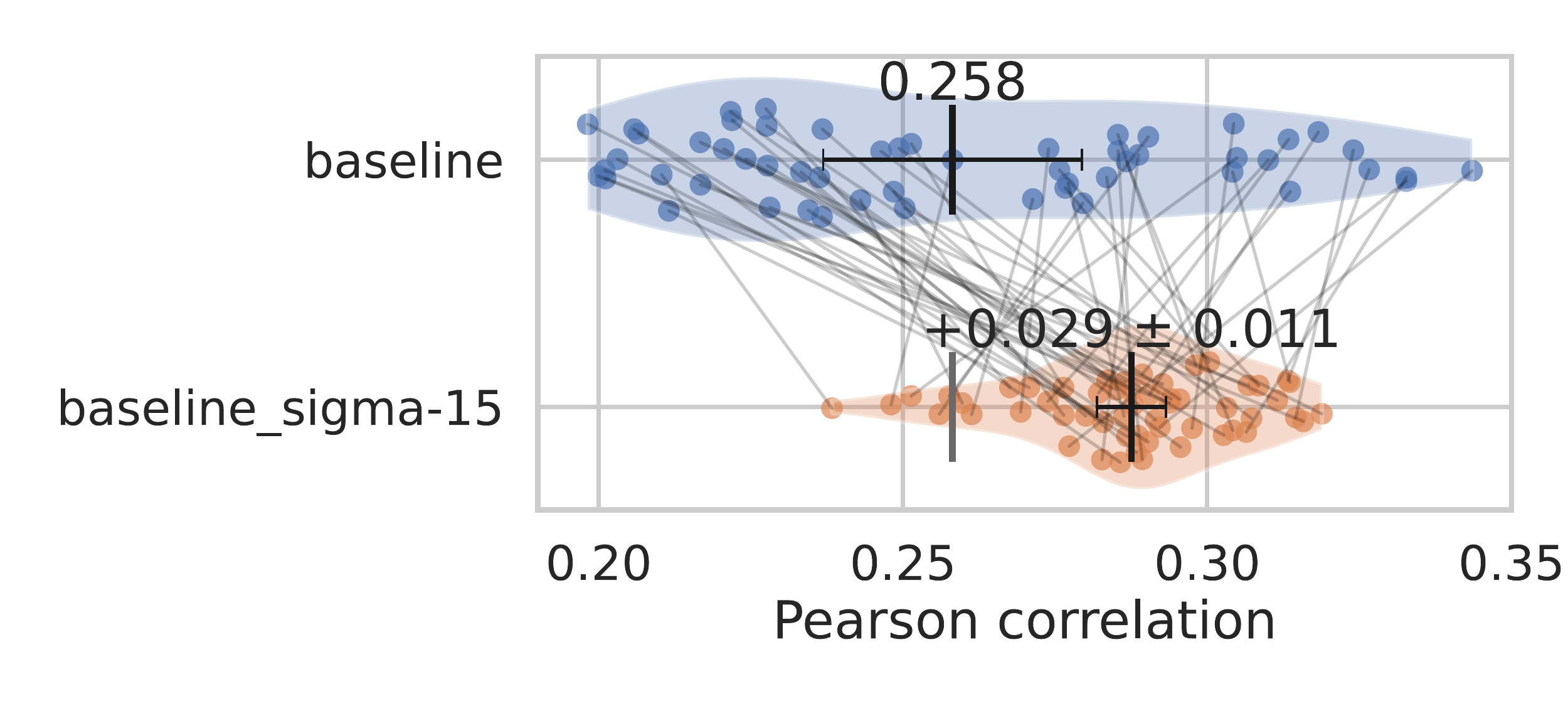}
    \includegraphics[width=0.49\columnwidth]{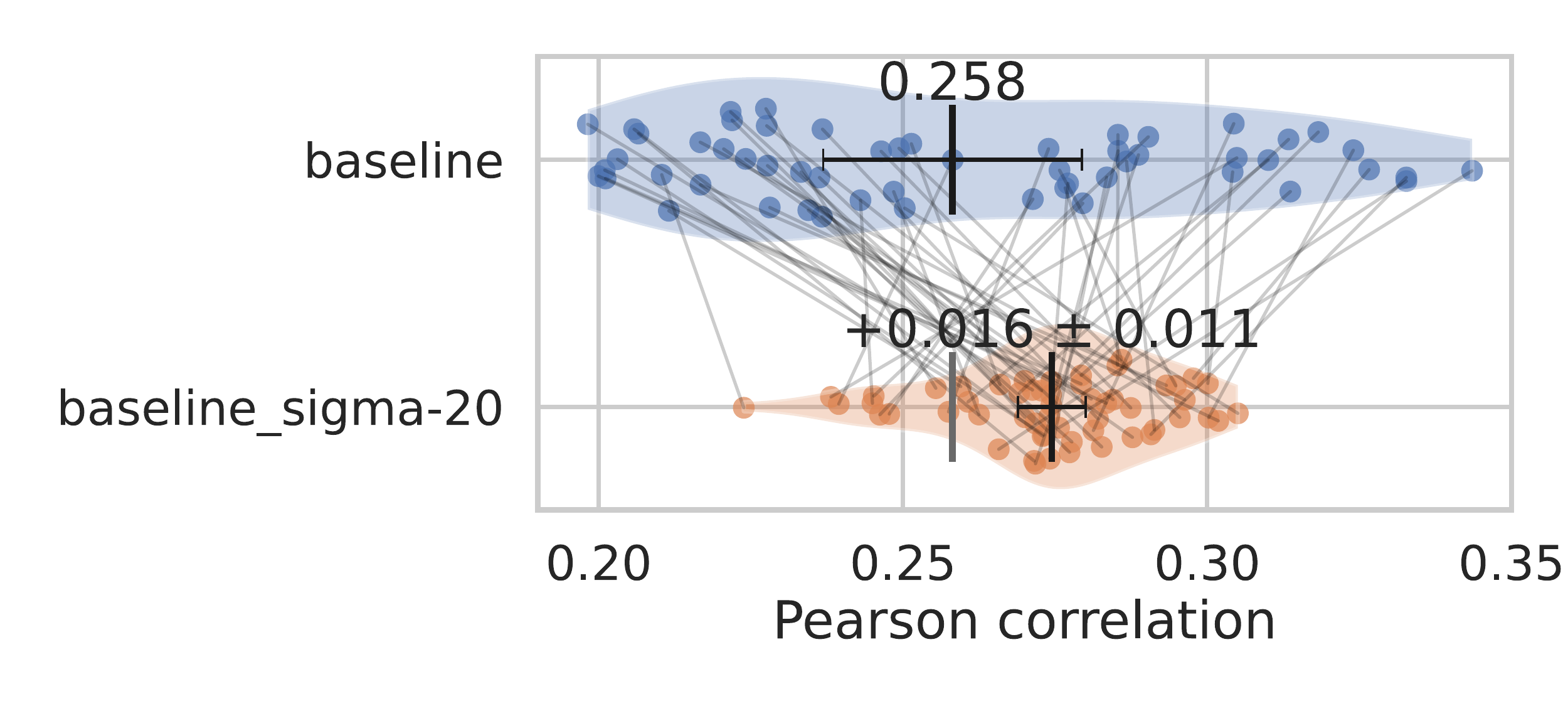}
    \caption{
        \textbf{Comparison of gains in correlation after Gaussian blurring}
        We compare correlation between subjects after the source subject's
        functional data has been smoothed with a Gaussian kernel of standard
        deviation 5mm (top left), 10mm (top right), 15mm (bottom left)
        and 20mm (bottom right)
    }
    \label{fig:baseline_smooth}
\end{figure}

Since smoothing signal on the source subject can reduce noise and increase correlation to target data,
we measure the correlation gain induced by applying a gaussian kernel to the source signal.
This allows us to show that only a minor proportion of correlation gains induced by FUGW can come from
this smoothing effect.
Figure \ref{fig:baseline_smooth} shows this for kernels of
5mm, 10mm, 15mm and 20mm of standard deviation respectively.
We see that correlation increases significantly less than when using FUGW
(0.03 vs 0.12 correlation gain respectively).
Moreover, one notices that even though correlation increases for pairs of subjects
with a low initial correlation, it decreases for pairs with a high initial correlation.
On the contrary, FUGW increases correlation for all pairs of subjects.

\paragraph{Different training sets yield comparable correlation gains}
While we leverage all IBC maps to derive our couplings, we show that the presented results hold when using a much smaller training dataset. In particular,
we observe similar correlation gains when using only the 57 maps of the Archi protocol for training (see Table \ref{tab:training_contrasts}). It takes about one hour per subject to acquire these maps, which we advocate is a reasonable amount of time to build a training set dedicated to align subjects within a given cohort (and possibly across cohorts).
Finally, we train both FUGW and MSM with lower-dimensional versions of the previous datasets. To do so, given a pair of subjects $(s, t)$ to be aligned, we fit a PCA on the left out subjects, project the data of subjects to be aligned on these components, and keep the first 20 components only. For both models, correlation gains remained unchanged.

More explicitly, we test the 4 following training sets:
\begin{itemize}
    \item ALL-MATH: all contrast maps of IBC except contrasts from the
    Mathlang protocol (369 features per subject)
    \item ALL-MATH PCA: principal components fitted on ALL-MATH
    for all IBC subjects except $s$ and $t$ (20 features)
    \item ARCHI: all contrast maps from the Archi protocol of IBC
    (57 features)
    \item ARCHI PCA: principal components fitted on ARCHI
    for all subjects except $s$ and $t$ (20 features)
\end{itemize}

\begin{table}[ht]
    \begin{center}
    \begin{tabular}{l|ll}
        Training set    & FUGW      & MSM   \\ \hline
        ALL-MATH        & 0.12      & 0.01     \\
        ALL-MATH PCA    & 0.11      & 0.02     \\
        ARCHI           & 0.10      & 0.02     \\
        ARCHI PCA       & 0.11      & 0.01
    \end{tabular}
    \end{center}

    \caption{
        \textbf{Gain in Pearson correlation of aligned contrast maps
        from the Mathlang protocol compared to the baseline}
        The original correlation (baseline) is 0.258
    }
    \label{tab:varying_training_sets}
\end{table}

\paragraph{Using naturalistic stimuli to derive alignments with FUGW}

This experiment's setup is similar to that of Experiment 1: Using training features to first derive OT couplings, we then use the latter to assess correlation gains between subjects's feature maps before and after subjects have been aligned.
Naturalistic stimuli datasets include \emph{Raiders of the Lost Ark}, short video clips and auditory stimuli from \emph{The Little Prince} respectively adapted from  \cite{haxby_common_2011,nishimoto2011reconstructing,bhattasali2019localising}.
Here, for each naturalistic dataset, we leverage work from \cite{richard2020modeling} to derive the first $m=20$ components of a fitted shared response model. Share response models seek to find a common dictionary $\bm{K}$
of activation patterns across subjects $s \in S$
and to derive a mapping $\bm{W^s}$ with $m$ orthogonal components
that projects each individual's data onto this common space:
\begin{equation*}
    \label{eq:srm}
    \operatorname*{arg\,min}_{\substack{\bm{K}, \bm{W^s}\\\text{s.t. } \bm{(W^s)^T \cdot W^s = I_m}}}
    \sum_{s \in S} ||  \bm{F^s} - \bm{W^s} \bm{K} ||^2 
\end{equation*}
These $\bm{W^s}$ are then used for alignment.

Results are reported in Table \ref{tab:naturalistic_results} and show that using datasets which are 20 times less time-consuming than that of Experiment 1 can already yield significant correlation gain on unseen task data.

\begin{table}[!ht]
    \begin{center}
        \caption{Acquisition time (AT) and correlation gain on the left hemisphere (CG) per training set (baseline correlation = 0.258)}
        \label{tab:naturalistic_results}
        \vskip 0.12in
        \begin{tabular}{l|lll}
            Training set        & Type &  AT (min) & CG (Pearson)\\
            \hline
            All-MATH            & tasks & 2000  & 0.118\\
            Clips               & movie & 100   & 0.017\\
            Raiders             & movie & 115   & 0.046\\
            The Little Prince   & movie & 100   & 0.009\\
        \end{tabular}
    \end{center}
\end{table}

\begin{figure}[ht]
    \centering
    \includegraphics[width=0.49\columnwidth]{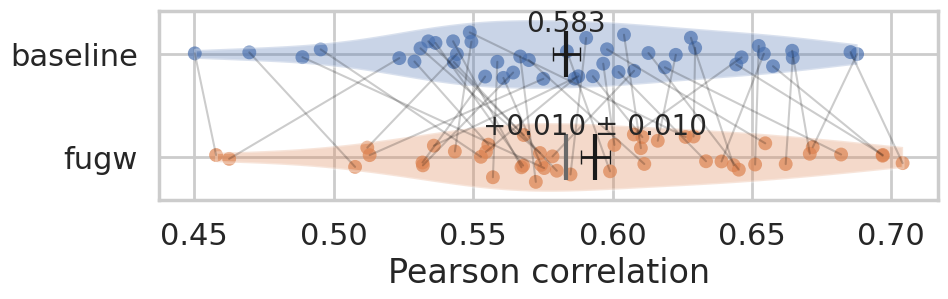}
    \includegraphics[width=0.49\columnwidth]{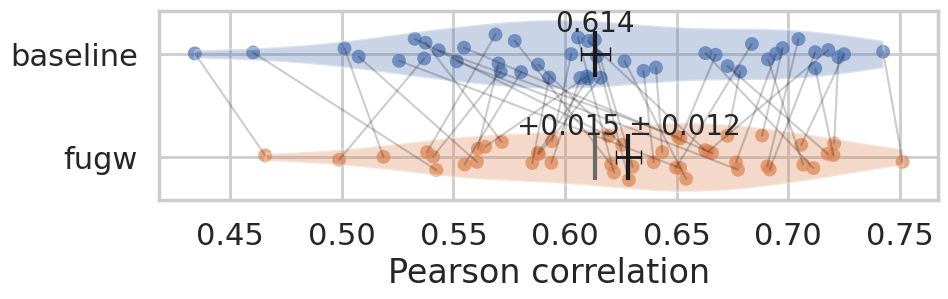}
    \caption{
        \textbf{Comparison of gains in correlation after inter-subject alignment for myelin maps}
        For each pair of source and target subjects
        of the IBC dataset, we compute the average Pearson correlation between
        myelin maps -- approximated using T1/T2 ratios --
        for the left (left panel) and right (right panel) hemispheres.
        Correlation gains are not significant.
    }
    \label{fig:myelin_correlation_gains}
\end{figure}

\paragraph{Transporting myelin maps shows mild effect}
Leveraging transport plans computed using fMRI data from Experiment 1, we transport myelin maps -- approximated through T1 / T2 ratio maps -- from the source subject to the target subject. We compare the correlation of the unaligned source and target maps with the correlation of the transported and target maps. As illustrated in Figure \ref{fig:myelin_correlation_gains}, correlation gain is barely significant. 

Before computing correlation between aforementioned maps, we discarded vertices located in the \textit{cortical wall}, as they mostly contain spurious values. To do so, we they their value to the median of values of vertices which do not belong to the cortical wall. In order to determine which vertices belong to the \textit{wall}, we used the Destrieux atlas \cite{destrieux_automatic_2010}. 

\begin{figure}[H]
    \centering
    \includegraphics[width=1\columnwidth]{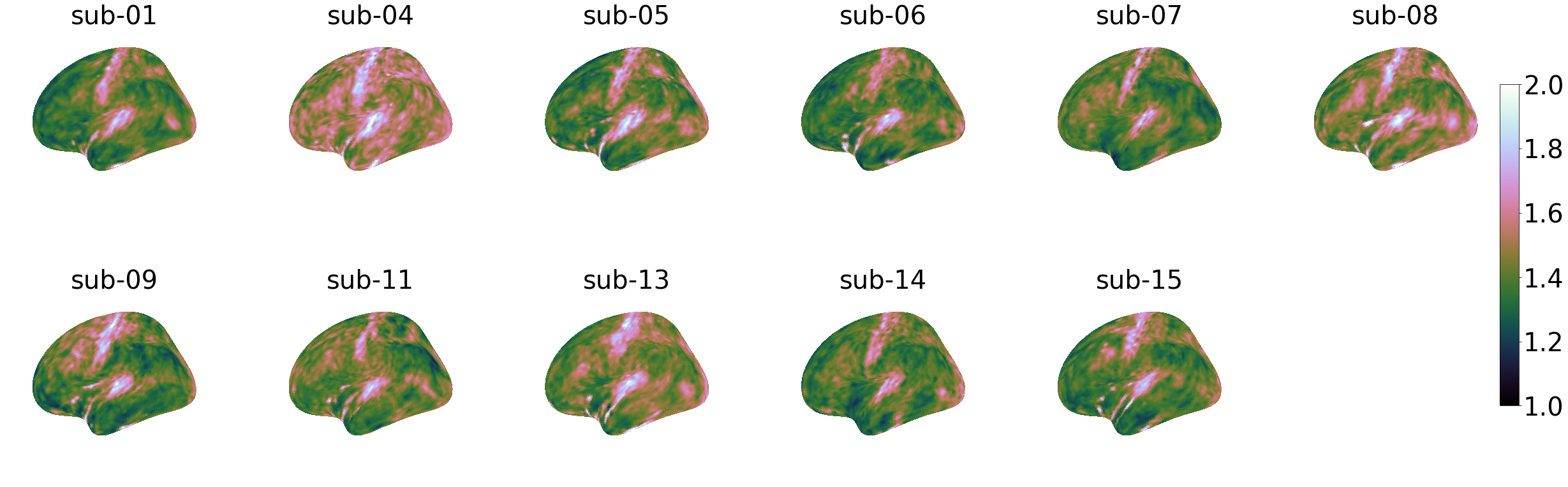}
    \caption{
        \textbf{Myelin maps (approximated using T1/T2 ratio maps) are very consistent across IBC participants}
    }
    \label{fig:ibc_myelin_maps}
\end{figure}

Eventually, we advocate that little gain can be obtained when better aligning myelin maps, since they are already very stable across human subjects as shown in Figure \ref{fig:ibc_myelin_maps}.

\subsection{Dataset description}
The presented experiments rely on the Individual Brain Charting (IBC) dataset.
A detailed description of the preprocessing pipeline of the IBC data is provided
in \cite{Pinho2021}.
Raw data were preprocessed using \emph{PyPreprocess}
\footnote{\url{https://github.com/neurospin/pypreprocess}}.

All fMRI images, i.e. GE-EPI volumes, were collected twice with reversed
phase-encoding directions, resulting in pairs of images with distortions going
in opposite directions.
Susceptibility-induced off-resonance field was estimated from the two Spin-Echo
EPI volumes in reversed phase-encoding directions.
The images were corrected based on the estimated deformation model.
Details about the method can be found in \cite{Andersson2003}.

Further, the GE-EPI volumes were aligned to each other within every participant.
A rigid-body transformation was employed, in which the average volume of all
images was used as reference \cite{Friston1995}.
The anatomical and motion-corrected fMRI images were given as input to
\emph{FreeSurfer} v6.0.0, in order to extract meshes of the tissue
interfaces and the sampling of functional activation on these meshes, as
described in \cite{vanessen2012}.
The corresponding maps were then resampled to the fsaverage7 (high resolution, 163k nodes per hemisphere) and fsaverage5 (low resolution, 10k nodes per hemisphere) templates of
FreeSurfer \cite{Fischl1999}.

FMRI data were analyzed using the \textit{General Linear Model}.
Regressors of the model were designed to capture variations in BOLD response
strictly following stimulus timing specifications.
They were estimated through the convolution of boxcar functions, that represent
per-condition stimulus occurrences, with the canonical \textit{Hemodynamic
  Response Function} (HRF).
To build such models, paradigm descriptors grouped in triplets (i.e. onset time,
duration and trial type) according to BIDS Specification were determined from
the log files' registries generated by the stimulus-delivery software.
To account for small fluctuations in the latency of the HRF peak response,
additional regressors were computed based on the convolution of the same
task-conditions profile with the time derivative of the HRF.
Nuisance regressors were also added to the design matrix in order to minimize
the final residual error.
To remove signal variance associated with spurious effects arising from
movements, six temporal regressors were defined for the motion parameters.
Further, the first five principal components of the signal, extracted from
voxels showing the $5\%$ highest variance, were also regressed to capture
physiological noise \cite{Behzadi2007}.

In addition, a discrete-cosine basis was included for high-pass filtering
(\textit{cutoff}=$\frac{1}{128}\textrm{Hz}$).
Model specification was implemented using \textit{Nilearn} v0.8.1
\cite{Abraham2014}, a Python library for statistical learning on neuroimaging
data (\url{https://nilearn.github.io}).

In tables \ref{tab:training_contrasts}, \ref{tab:validation_data} and \ref{tab:test_contrasts},
we give the explicit list of contrast and condition maps used for training, validation and testing respectively.

\newpage
\tablefirsthead{
    \multicolumn{2}{c}
    {{\bfseries Training data}}\\
    \toprule
    Task&\multicolumn{1}{l}{Condition / Contrast}\\
    \midrule
}
\tablehead{
    \multicolumn{2}{c}
    {{\bfseries Training data (next)}}\\
    \toprule
    Task&\multicolumn{1}{l}{Condition / Contrast}\\
    \midrule
}
\tablelasttail{
    \bottomrule
}
\topcaption{
    \textbf{Training data\label{tab:training_contrasts}}
    This dataset comprises a wide variety of tasks:
    motor, visual, auditory, relational, linguistic, etc.
}
\begin{supertabular}{ll}
ArchiEmotional & expression\_control\\
ArchiEmotional & expression\_gender\\
ArchiEmotional & expression\_gender-control\\
ArchiEmotional & expression\_intention\\
ArchiEmotional & expression\_intention-control\\
ArchiEmotional & expression\_intention-gender\\
ArchiEmotional & face\_control\\
ArchiEmotional & face\_gender\\
ArchiEmotional & face\_gender-control\\
ArchiEmotional & face\_trusty\\
ArchiEmotional & face\_trusty-control\\
ArchiEmotional & face\_trusty-gender\\
ArchiEmotional & trusty\_and\_intention-control\\
ArchiEmotional & trusty\_and\_intention-gender\\
ArchiSocial & false\_belief-mechanistic\\
ArchiSocial & false\_belief-mechanistic\_audio\\
ArchiSocial & false\_belief-mechanistic\_video\\
ArchiSocial & false\_belief\_audio\\
ArchiSocial & false\_belief\_video\\
ArchiSocial & mechanistic\_audio\\
ArchiSocial & mechanistic\_video\\
ArchiSocial & non\_speech\_sound\\
ArchiSocial & speech-non\_speech\\
ArchiSocial & speech\_sound\\
ArchiSocial & triangle\_mental\\
ArchiSocial & triangle\_mental-random\\
ArchiSocial & triangle\_random\\
ArchiSpatial & grasp-orientation\\
ArchiSpatial & hand-side\\
ArchiSpatial & object\_grasp\\
ArchiSpatial & object\_orientation\\
ArchiSpatial & rotation\_hand\\
ArchiSpatial & rotation\_side\\
ArchiSpatial & saccades\\
ArchiStandard & audio\_computation\\
ArchiStandard & audio\_left\_button\_press\\
ArchiStandard & audio\_right\_button\_press\\
ArchiStandard & audio\_sentence\\
ArchiStandard & cognitive-motor\\
ArchiStandard & computation\\
ArchiStandard & computation-sentences\\
ArchiStandard & horizontal-vertical\\
ArchiStandard & horizontal\_checkerboard\\
ArchiStandard & left-right\_button\_press\\
ArchiStandard & listening-reading\\
ArchiStandard & motor-cognitive\\
ArchiStandard & reading-checkerboard\\
ArchiStandard & reading-listening\\
ArchiStandard & right-left\_button\_press\\
ArchiStandard & sentences\\
ArchiStandard & sentences-computation\\
ArchiStandard & vertical-horizontal\\
ArchiStandard & vertical\_checkerboard\\
ArchiStandard & video\_computation\\
ArchiStandard & video\_left\_button\_press\\
ArchiStandard & video\_right\_button\_press\\
ArchiStandard & video\_sentence\\
Attention & double\_congruent\\
Attention & double\_cue\\
Attention & double\_incongruent\\
Attention & double\_incongruent-double\_congruent\\
Attention & incongruent-congruent\\
Attention & spatial\_congruent\\
Attention & spatial\_cue\\
Attention & spatial\_cue-double\_cue\\
Attention & spatial\_incongruent\\
Attention & spatial\_incongruent-spatial\_congruent\\
Audi & alphabet\\
Audi & alphabet-silence\\
Audi & animals\\
Audi & animals-silence\\
Audi & cough\\
Audi & cough-silence\\
Audi & environment\\
Audi & environment-silence\\
Audi & human\\
Audi & human-silence\\
Audi & laugh\\
Audi & laugh-silence\\
Audi & music\\
Audi & music-silence\\
Audi & reverse\\
Audi & reverse-silence\\
Audi & silence\\
Audi & speech\\
Audi & speech-silence\\
Audi & suomi\\
Audi & suomi-silence\\
Audi & tear\\
Audi & tear-silence\\
Audi & yawn\\
Audi & yawn-silence\\
Audio & animal\\
Audio & animal-others\\
Audio & animal-silence\\
Audio & mean-silence\\
Audio & music\\
Audio & music-others\\
Audio & music-silence\\
Audio & nature\\
Audio & nature-others\\
Audio & nature-silence\\
Audio & speech\\
Audio & speech-others\\
Audio & speech-silence\\
Audio & tool\\
Audio & tool-others\\
Audio & tool-silence\\
Audio & voice\\
Audio & voice-others\\
Audio & voice-silence\\
Bang & no\_talk\\
Bang & talk\\
Bang & talk-no\_talk\\
ColumbiaCards & gain\\
ColumbiaCards & loss\\
ColumbiaCards & num\_loss\_cards\\
Discount & amount\\
Discount & delay\\
DotPatterns & correct\_cue-incorrect\_cue\\
DotPatterns & correct\_cue\_correct\_probe\\
DotPatterns & correct\_cue\_incorrect\_probe\\
DotPatterns & correct\_cue\_incorrect\_probe-correct\_cue\_correct\_probe\\
DotPatterns & correct\_cue\_incorrect\_probe-incorrect\_cue\_correct\_probe\\
DotPatterns & cue\\
DotPatterns & incorrect\_cue\_correct\_probe\\
DotPatterns & incorrect\_cue\_incorrect\_probe\\
DotPatterns & incorrect\_cue\_incorrect\_probe-correct\_cue\_incorrect\_probe\\
DotPatterns & incorrect\_cue\_incorrect\_probe-incorrect\_cue\_correct\_probe\\
DotPatterns & incorrect\_probe-correct\_probe\\
EmotionalPain & emotional-physical\_pain\\
EmotionalPain & emotional\_pain\\
EmotionalPain & physical\_pain\\
Enumeration & enumeration\_constant\\
Enumeration & enumeration\_linear\\
Enumeration & enumeration\_quadratic\\
Lec1 & pseudoword\\
Lec1 & pseudoword-random\_string\\
Lec1 & random\_string\\
Lec1 & word\\
Lec1 & word-pseudoword\\
Lec1 & word-random\_string\\
Lec2 & attend\\
Lec2 & attend-unattend\\
Lec2 & unattend\\
MCSE & high-low\_salience\\
MCSE & high\_salience\_left\\
MCSE & high\_salience\_right\\
MCSE & low+high\_salience\\
MCSE & low-high\_salience\\
MCSE & low\_salience\_left\\
MCSE & low\_salience\_right\\
MCSE & salience\_left-right\\
MCSE & salience\_right-left\\
MTTNS & northside-southside\_event\\
MTTNS & sn\_after-before\_event\\
MTTNS & sn\_after\_event\\
MTTNS & sn\_all\_event\_response\\
MTTNS & sn\_all\_space-time\_cue\\
MTTNS & sn\_all\_space\_cue\\
MTTNS & sn\_all\_time-space\_cue\\
MTTNS & sn\_all\_time\_cue\\
MTTNS & sn\_average\_event\\
MTTNS & sn\_average\_reference\\
MTTNS & sn\_before-after\_event\\
MTTNS & sn\_before\_event\\
MTTNS & sn\_northside\_event\\
MTTNS & sn\_southside\_event\\
MTTNS & sn\_space-time\_event\\
MTTNS & sn\_space\_event\\
MTTNS & sn\_time-space\_event\\
MTTNS & sn\_time\_event\\
MTTNS & southside-northside\_event\\
MTTWE & eastside-westside\_event\\
MTTWE & we\_after-before\_event\\
MTTWE & we\_after\_event\\
MTTWE & we\_all\_event\_response\\
MTTWE & we\_all\_space-time\_cue\\
MTTWE & we\_all\_space\_cue\\
MTTWE & we\_all\_time-space\_cue\\
MTTWE & we\_all\_time\_cue\\
MTTWE & we\_average\_event\\
MTTWE & we\_average\_reference\\
MTTWE & we\_before-after\_event\\
MTTWE & we\_before\_event\\
MTTWE & we\_eastside\_event\\
MTTWE & we\_space-time\_event\\
MTTWE & we\_space\_event\\
MTTWE & we\_time-space\_event\\
MTTWE & we\_time\_event\\
MTTWE & we\_westside\_event\\
MTTWE & westside-eastside\_event\\
MVEB & 2\_letters\_different\\
MVEB & 2\_letters\_different-same\\
MVEB & 2\_letters\_same\\
MVEB & 4\_letters\_different\\
MVEB & 4\_letters\_different-same\\
MVEB & 4\_letters\_same\\
MVEB & 6\_letters\_different\\
MVEB & 6\_letters\_different-2\_letters\_different\\
MVEB & 6\_letters\_different-same\\
MVEB & 6\_letters\_same\\
MVEB & letter\_occurrence\_response\\
MVIS & 2\_dots-2\_dots\_control\\
MVIS & 4\_dots-4\_dots\_control\\
MVIS & 6\_dots-2\_dots\\
MVIS & 6\_dots-6\_dots\_control\\
MVIS & dot\_displacement\_response\\
MVIS & dots-control\\
Moto & finger\_left-fixation\\
Moto & finger\_right-fixation\\
Moto & foot\_left-fixation\\
Moto & foot\_right-fixation\\
Moto & hand\_left-fixation\\
Moto & hand\_right-fixation\\
Moto & instructions\\
Moto & saccade-fixation\\
Moto & tongue-fixation\\
PainMovie & movie\_mental\\
PainMovie & movie\_mental-pain\\
PainMovie & movie\_pain\\
Preference & face-others\\
Preference & food-others\\
Preference & house-others\\
Preference & painting-others\\
Preference & preference\_constant\\
Preference & preference\_linear\\
Preference & preference\_quadratic\\
PreferenceFaces & face\_constant\\
PreferenceFaces & face\_linear\\
PreferenceFaces & face\_quadratic\\
PreferenceFood & food\_constant\\
PreferenceFood & food\_linear\\
PreferenceFood & food\_quadratic\\
PreferenceHouses & house\_constant\\
PreferenceHouses & house\_linear\\
PreferenceHouses & house\_quadratic\\
PreferencePaintings & painting\_constant\\
PreferencePaintings & painting\_linear\\
PreferencePaintings & painting\_quadratic\\
RSVPLanguage & complex\\
RSVPLanguage & complex-consonant\_string\\
RSVPLanguage & complex-simple\\
RSVPLanguage & consonant\_string\\
RSVPLanguage & jabberwocky\\
RSVPLanguage & jabberwocky-consonant\_string\\
RSVPLanguage & jabberwocky-pseudo\\
RSVPLanguage & probe\\
RSVPLanguage & pseudo-consonant\_string\\
RSVPLanguage & pseudoword\_list\\
RSVPLanguage & sentence-consonant\_string\\
RSVPLanguage & sentence-jabberwocky\\
RSVPLanguage & sentence-pseudo\\
RSVPLanguage & sentence-word\\
RSVPLanguage & simple\\
RSVPLanguage & simple-consonant\_string\\
RSVPLanguage & word-consonant\_string\\
RSVPLanguage & word-pseudo\\
RSVPLanguage & word\_list\\
SelectiveStopSignal & go\_critical\\
SelectiveStopSignal & go\_critical-stop\\
SelectiveStopSignal & go\_noncritical\\
SelectiveStopSignal & go\_noncritical-ignore\\
SelectiveStopSignal & ignore\\
SelectiveStopSignal & ignore-stop\\
SelectiveStopSignal & stop\\
SelectiveStopSignal & stop-ignore\\
Self & correct\_rejection\\
Self & encode\_other\\
Self & encode\_self\\
Self & encode\_self-other\\
Self & false\_alarm\\
Self & instructions\\
Self & recognition\_hit\\
Self & recognition\_hit-correct\_rejection\\
Self & recognition\_other\_hit\\
Self & recognition\_self-other\\
Self & recognition\_self\_hit\\
StopSignal & go\\
StopSignal & stop\\
StopSignal & stop-go\\
Stroop & congruent\\
Stroop & incongruent\\
Stroop & incongruent-congruent\\
TheoryOfMind & belief\\
TheoryOfMind & belief-photo\\
TheoryOfMind & photo\\
TwoByTwo & cue\_switch-stay\\
TwoByTwo & cue\_taskstay\_cuestay\\
TwoByTwo & cue\_taskstay\_cueswitch\\
TwoByTwo & cue\_taskswitch\_cuestay\\
TwoByTwo & cue\_taskswitch\_cueswitch\\
TwoByTwo & stim\_taskstay\_cuestay\\
TwoByTwo & stim\_taskstay\_cueswitch\\
TwoByTwo & stim\_taskswitch\_cuestay\\
TwoByTwo & stim\_taskswitch\_cueswitch\\
TwoByTwo & task\_switch-stay\\
VSTM & vstm\_constant\\
VSTM & vstm\_linear\\
VSTM & vstm\_quadratic\\
Visu & animal\\
Visu & animal-scrambled\\
Visu & characters\\
Visu & characters-scrambled\\
Visu & face\\
Visu & face-scrambled\\
Visu & house\\
Visu & house-scrambled\\
Visu & pseudoword\\
Visu & pseudoword-scrambled\\
Visu & scene\\
Visu & scene-scrambled\\
Visu & scrambled\\
Visu & target\_fruit\\
Visu & tool\\
Visu & tool-scrambled\\
WardAndAllport & ambiguous-unambiguous\\
WardAndAllport & intermediate-direct\\
WardAndAllport & move\_ambiguous\_direct\\
WardAndAllport & move\_ambiguous\_intermediate\\
WardAndAllport & move\_unambiguous\_direct\\
WardAndAllport & move\_unambiguous\_intermediate\\
WardAndAllport & planning\_ambiguous\_direct\\
WardAndAllport & planning\_ambiguous\_intermediate\\
WardAndAllport & planning\_unambiguous\_direct\\
WardAndAllport & planning\_unambiguous\_intermediate\\



\end{supertabular}


\newpage
\tablefirsthead{
    \multicolumn{2}{c}
    {{\bfseries Validation data}}\\
    \toprule
    Task&\multicolumn{1}{c}{Condition / Contrast} \\
    \midrule
}
\tablehead{
    \multicolumn{2}{c}
    {{\bfseries Validation data (next)}} \\
    \toprule
    Task&\multicolumn{1}{c}{Condition / Contrast}\\
    \midrule
}
\tablelasttail{
    \bottomrule
}
\topcaption{Validation data\label{tab:validation_data}}
\begin{supertabular}{ll}
HcpEmotion & face\\
HcpEmotion & face-shape\\
HcpEmotion & shape\\
HcpEmotion & shape-face\\
HcpGambling & punishment\\
HcpGambling & punishment-reward\\
HcpGambling & reward\\
HcpGambling & reward-punishment\\
HcpLanguage & math\\
HcpLanguage & math-story\\
HcpLanguage & story\\
HcpLanguage & story-math\\
HcpMotor & cue\\
HcpMotor & left\_foot\\
HcpMotor & left\_foot-avg\\
HcpMotor & left\_hand\\
HcpMotor & left\_hand-avg\\
HcpMotor & right\_foot\\
HcpMotor & right\_foot-avg\\
HcpMotor & right\_hand\\
HcpMotor & right\_hand-avg\\
HcpMotor & tongue\\
HcpMotor & tongue-avg\\
HcpRelational & match\\
HcpRelational & relational\\
HcpRelational & relational-match\\
HcpSocial & mental\\
HcpSocial & mental-random\\
HcpSocial & random\\
HcpWm & 0back-2back\\
HcpWm & 0back\_body\\
HcpWm & 0back\_face\\
HcpWm & 0back\_place\\
HcpWm & 0back\_tools\\
HcpWm & 2back-0back\\
HcpWm & 2back\_body\\
HcpWm & 2back\_face\\
HcpWm & 2back\_place\\
HcpWm & 2back\_tools\\
HcpWm & body-avg\\
HcpWm & face-avg\\
HcpWm & place-avg\\
HcpWm & tools-avg\\
\end{supertabular}

\newpage
\tablefirsthead{
    \multicolumn{2}{c}
    {{\bfseries Test data}} \\
    \toprule
    Task&\multicolumn{1}{c}{Condition / Contrast} \\
    \midrule
}
\tablehead{
    \multicolumn{2}{c}
    {{\bfseries Test data (next)}} \\
    \toprule
    Task&\multicolumn{1}{c}{Condition / Contrast}\\
    \midrule
}
\tablelasttail{
    \bottomrule
}
\topcaption{Test data\label{tab:test_contrasts}}
\begin{supertabular}{ll}
MathLanguage & arithmetic\_fact-othermath\\
MathLanguage & arithmetic\_fact\_auditory\\
MathLanguage & arithmetic\_fact\_visual\\
MathLanguage & arithmetic\_principle-othermath\\
MathLanguage & arithmetic\_principle\_auditory\\
MathLanguage & arithmetic\_principle\_visual\\
MathLanguage & auditory-visual\\
MathLanguage & colorlessg-wordlist\\
MathLanguage & colorlessg\_auditory\\
MathLanguage & colorlessg\_visual\\
MathLanguage & context-general\\
MathLanguage & context-theory\_of\_mind\\
MathLanguage & context\_auditory\\
MathLanguage & context\_visual\\
MathLanguage & general-colorlessg\\
MathLanguage & general\_auditory\\
MathLanguage & general\_visual\\
MathLanguage & geometry-othermath\\
MathLanguage & geometry\_fact\_auditory\\
MathLanguage & geometry\_fact\_visual\\
MathLanguage & math-nonmath\\
MathLanguage & nonmath-math\\
MathLanguage & theory\_of\_mind-context\\
MathLanguage & theory\_of\_mind-general\\
MathLanguage & theory\_of\_mind\_and\_context-general\\
MathLanguage & theory\_of\_mind\_auditory\\
MathLanguage & theory\_of\_mind\_visual\\
MathLanguage & visual-auditory\\
MathLanguage & wordlist\_auditory\\
MathLanguage & wordlist\_visual\\
\end{supertabular}

\newpage
\printbibliography

@article{Abraham2014,
AUTHOR={Abraham, Alexandre and Pedregosa, Fabian and Eickenberg, Michael and Gervais, Philippe and Mueller, Andreas and Kossaifi, Jean and Gramfort, Alexandre and Thirion, Bertrand and Varoquaux, Ga{\"e}},
TITLE={Machine learning for neuroimaging with scikit-learn},
JOURNAL={Front Neuroinform},
VOLUME={8},
PAGES={14},
YEAR={2014},
URL={https://doi.org/10.3389/fninf.2014.00014},
ISSN={1662-5196},
}

@article{Andersson2003,
title = "How to correct susceptibility distortions in spin-echo echo-planar images: application to diffusion tensor imaging",
journal = "Neuroimage",
volume = "20",
number = "2",
pages = "870 - 888",
year = "2003",
issn = "1053-8119",
url = "http://doi.org/10.1016/S1053-8119(03)00336-7",
author = "Jesper L.R. Andersson and Stefan Skare and John Ashburner"
}

@article{Behzadi2007,
title = "A component based noise correction method (CompCor) for \{BOLD\} and perfusion based f{MRI}",
journal = "Neuroimage",
volume = "37",
number = "1",
pages = "90 - 101",
year = "2007",
issn = "1053-8119",
url = "https://doi.org/10.1016/j.neuroimage.2007.04.042",
author = "Yashar Behzadi and Khaled Restom and Joy Liau and Thomas T. Liu"
}

@article{Fischl1999,
author = {Fischl, Bruce and Sereno, Martin I. and Tootell, Roger B.H. and Dale, Anders M.},
title = {High-resolution intersubject averaging and a coordinate system for the cortical surface},
journal = {Hum Brain Mapp},
volume = {8},
number = {4},
pages = {272-284},
url = {https://doi.org/10.1002/(SICI)1097-0193(1999)8:4<272::AID-HBM10>3.0.CO;2-4},
year = {1999}
}

@article{Friston1995,
    author       = "K.J. Friston and C.D. Frith and R.S.J. Frackowiak and R. Turner",
    title        = "Characterizing {D}ynamic {B}rain {R}esponses with f{MRI}: a {M}ultivariate {A}pproach",
    journal      = "Neuroimage",
    volume       = "2",
    number       = "2",
    pages        = "166-172",
    year         = "1995",
    url          = "https://doi.org/10.1006/nimg.1995.1019",
}

@article{Pinho2021,
author = {Pinho, Ana Lu{\'i}sa and Amadon, Alexis and Fabre, Murielle and Dohmatob, Elvis and Denghien, Isabelle and Torre, Juan Jes{\'u}s and Ginisty, Chantal and Becuwe-Desmidt, S{\'e}verine and Roger, S{\'e}verine and Laurier, Laurence and Joly-Testault, V{\'e}ronique and M{\'e}diouni-Cloarec, Ga{\"e}lle and Doubl{\'e}, Christine and Martins, Bernadette and Pinel, Philippe and Eger, Evelyn and Varoquaux, Ga{\"e}l and Pallier, Christophe and Dehaene, Stanislas and Hertz-Pannier, Lucie and Thirion, Bertrand},
title = {Subject-specific segregation of functional territories based on deep phenotyping},
journal = {Hum Brain Mapp},
volume = {42},
number = {4},
pages = {841-870},
year = {2021},
url = {https://doi.org/10.1002/hbm.25189},
annotate = {* This paper introduces several key experiments on a fraction of the IBC dataset, namely the first release. In particular, it introduces the application of dictionary learning to summarize contrast maps to topographies. It studies the stability of the dictionary components across data resamplings. It also shows that some contrast maps can be successfully reconstructed from other contrasts. Finally, it illustrates how the accumulation of functional contrasts can help to distinguish between the functional specialization of several regions taken from the language network.}
}

@article{neubert_comparison_2014,
	title = {Comparison of human ventral frontal cortex areas for cognitive control and language with areas in monkey frontal cortex},
	volume = {81},
	issn = {1097-4199},
	doi = {10.1016/j.neuron.2013.11.012},
	pages = {700--713},
	number = {3},
	journaltitle = {Neuron},
	shortjournal = {Neuron},
	author = {Neubert, Franz-Xaver and Mars, Rogier B. and Thomas, Adam G. and Sallet, Jerome and Rushworth, Matthew F. S.},
	date = {2014-02-05},
	pmid = {24485097},
}

@inproceedings{bazeille_local_2019,
	location = {Hong Kong, China},
	title = {Local Optimal Transport for Functional Brain Template Estimation},
	url = {https://hal.archives-ouvertes.fr/hal-02278663},
	doi = {10.1007/978-3-030-20351-1_18},
	abstract = {An important goal of cognitive brain imaging studies is to model the functional organization of the brain; yet there exists currently no functional brain atlas built from existing data. One of the main roadblocks to the creation of such an atlas is the functional variability that is observed in subjects performing the same task; this variability goes far beyond anatomical variability in brain shape and size. Function-based alignment procedures have recently been proposed in order to improve the correspondence of activation patterns across individuals. However, the corresponding computational solutions are costly and not well-principled. Here, we propose a new framework based on optimal transport theory to create such a template. We leverage entropic smoothing as an efficient means to create brain templates without losing fine-grain structural information; it is implemented in a computationally efficient way. We evaluate our approach on rich multi-subject, multi-contrasts datasets. These experiments demonstrate that the template-based inference procedure improves the transfer of information across individuals with respect to state of the art methods.},
	booktitle = {{IPMI} 2019 - 26th International Conference on Information Processing in Medical Imaging},
	author = {{BAZEILLE}, Thomas and Richard, Hugo and Janati, Hicham and Thirion, Bertrand},
	% %urldate = {2020-10-27},
	date = {2019-06},
	keywords = {Brain, Atlas Inference, {FMRI}, Functionnal Alignment, Read, Printed},
	file = {HAL PDF Full Text:/home/alexis/snap/zotero-snap/common/Zotero/storage/BJBMZV7F/BAZEILLE et al. - 2019 - Local Optimal Transport for Functional Brain Templ.pdf:application/pdf},
}

@article{mars_whole_2018,
	title = {Whole brain comparative anatomy using connectivity blueprints},
	volume = {7},
	issn = {2050-084X},
	url = {https://doi.org/10.7554/eLife.35237},
	doi = {10.7554/eLife.35237},
	abstract = {Comparing the brains of related species faces the challenges of establishing homologies whilst accommodating evolutionary specializations. Here we propose a general framework for understanding similarities and differences between the brains of primates. The approach uses white matter blueprints of the whole cortex based on a set of white matter tracts that can be anatomically matched across species. The blueprints provide a common reference space that allows us to navigate between brains of different species, identify homologous cortical areas, or to transform whole cortical maps from one species to the other. Specializations are cast within this framework as deviations between the species’ blueprints. We illustrate how this approach can be used to compare human and macaque brains.},
	pages = {e35237},
	journaltitle = {{eLife}},
	author = {Mars, Rogier B and Sotiropoulos, Stamatios N and Passingham, Richard E and Sallet, Jerome and Verhagen, Lennart and Khrapitchev, Alexandre A and Sibson, Nicola and Jbabdi, Saad},
	editor = {Stephan, Klaas Enno},
	% urldate = {2020-09-30},
	date = {2018-05-11},
	note = {Publisher: {eLife} Sciences Publications, Ltd},
	keywords = {connectivity, comparative anatomy, connectivity blueprint, connectivity fingerprint matching, Read, Printed},
	file = {Full Text PDF:/home/alexis/snap/zotero-snap/common/Zotero/storage/47UTXJ5K/Mars et al. - 2018 - Whole brain comparative anatomy using connectivity.pdf:application/pdf},
}

@article{xu_cross-species_2020,
	title = {Cross-species functional alignment reveals evolutionary hierarchy within the connectome},
	volume = {223},
	issn = {1053-8119},
	url = {http://www.sciencedirect.com/science/article/pii/S1053811920308326},
	doi = {10.1016/j.neuroimage.2020.117346},
	abstract = {Evolution provides an important window into how cortical organization shapes function and vice versa. The complex mosaic of changes in brain morphology and functional organization that have shaped the mammalian cortex during evolution, complicates attempts to chart cortical differences across species. It limits our ability to fully appreciate how evolution has shaped our brain, especially in systems associated with unique human cognitive capabilities that lack anatomical homologues in other species. Here, we develop a function-based method for cross-species alignment that enables the quantification of homologous regions between humans and rhesus macaques, even when their location is decoupled from anatomical landmarks. Critically, we find cross-species similarity in functional organization reflects a gradient of evolutionary change that decreases from unimodal systems and culminates with the most pronounced changes in posterior regions of the default mode network (angular gyrus, posterior cingulate and middle temporal cortices). Our findings suggest that the establishment of the default mode network, as the apex of a cognitive hierarchy, has changed in a complex manner during human evolution – even within subnetworks.},
	pages = {117346},
	journaltitle = {{NeuroImage}},
	shortjournal = {{NeuroImage}},
	author = {Xu, Ting and Nenning, Karl-Heinz and Schwartz, Ernst and Hong, Seok-Jun and Vogelstein, Joshua T. and Goulas, Alexandros and Fair, Damien A. and Schroeder, Charles E. and Margulies, Daniel S. and Smallwood, Jonny and Milham, Michael P. and Langs, Georg},
	% urldate = {2020-09-22},
	date = {2020-12-01},
	langid = {english},
	keywords = {Evolution, Cross-species alignment, Default mode network, Hierarchy, Joint embedding},
	file = {Submitted Version:/home/alexis/snap/zotero-snap/common/Zotero/storage/S84N64TH/Xu et al. - 2020 - Cross-species functional alignment reveals evoluti.pdf:application/pdf},
}

@article{eichert_cross-species_2020,
	title = {Cross-species cortical alignment identifies different types of anatomical reorganization in the primate temporal lobe},
	volume = {9},
	issn = {2050-084X},
	url = {https://doi.org/10.7554/eLife.53232},
	doi = {10.7554/eLife.53232},
	abstract = {Evolutionary adaptations of temporo-parietal cortex are considered to be a critical specialization of the human brain. Cortical adaptations, however, can affect different aspects of brain architecture, including local expansion of the cortical sheet or changes in connectivity between cortical areas. We distinguish different types of changes in brain architecture using a computational neuroanatomy approach. We investigate the extent to which between-species alignment, based on cortical myelin, can predict changes in connectivity patterns across macaque, chimpanzee, and human. We show that expansion and relocation of brain areas can predict terminations of several white matter tracts in temporo-parietal cortex, including the middle and superior longitudinal fasciculus, but not the arcuate fasciculus. This demonstrates that the arcuate fasciculus underwent additional evolutionary modifications affecting the temporal lobe connectivity pattern. This approach can flexibly be extended to include other features of cortical organization and other species, allowing direct tests of comparative hypotheses of brain organization.},
	pages = {e53232},
	journaltitle = {{eLife}},
	author = {Eichert, Nicole and Robinson, Emma C and Bryant, Katherine L and Jbabdi, Saad and Jenkinson, Mark and Li, Longchuan and Krug, Kristine and Watkins, Kate E and Mars, Rogier B},
	editor = {Verstynen, Timothy and Gold, Joshua I and Verstynen, Timothy and Heuer, Katja},
	% urldate = {2020-08-27},
	date = {2020-03-23},
	note = {Publisher: {eLife} Sciences Publications, Ltd},
	keywords = {connectivity, Chimpanzee, cortical myelin, cross-species registration, temporal lobe, tractography, Read, To read again, Printed, Interesting},
	file = {Full Text PDF:/home/alexis/snap/zotero-snap/common/Zotero/storage/TNT55FWE/Eichert et al. - 2020 - Cross-species cortical alignment identifies differ.pdf:application/pdf},
}

@article{guntupalli_model_2016,
	title = {A Model of Representational Spaces in Human Cortex},
	volume = {26},
	issn = {1047-3211},
	url = {https://academic.oup.com/cercor/article/26/6/2919/1754308},
	doi = {10.1093/cercor/bhw068},
	abstract = {Abstract.  Current models of the functional architecture of human cortex emphasize areas that capture coarse-scale features of cortical topography but provide n},
	pages = {2919--2934},
	number = {6},
	journaltitle = {Cerebral Cortex},
	shortjournal = {Cereb Cortex},
	author = {Guntupalli, J. Swaroop and Hanke, Michael and Halchenko, Yaroslav O. and Connolly, Andrew C. and Ramadge, Peter J. and Haxby, James V.},
	%urldate = {2020-07-31},
	date = {2016-06-01},
	langid = {english},
	note = {Publisher: Oxford Academic},
	file = {Full Text PDF:/home/alexis/snap/zotero-snap/common/Zotero/storage/MAKFDBI3/Guntupalli et al. - 2016 - A Model of Representational Spaces in Human Cortex.pdf:application/pdf;Snapshot:/home/alexis/snap/zotero-snap/common/Zotero/storage/HBEM9SDR/1754308.html:text/html},
}

@article{robinson_multimodal_2018,
	title = {Multimodal surface matching with higher-order smoothness constraints},
	volume = {167},
	issn = {1095-9572},
	doi = {10.1016/j.neuroimage.2017.10.037},
	pages = {453--465},
	journaltitle = {{NeuroImage}},
	shortjournal = {Neuroimage},
	author = {Robinson, Emma C. and Garcia, Kara and Glasser, Matthew F. and Chen, Zhengdao and Coalson, Timothy S. and Makropoulos, Antonios and Bozek, Jelena and Wright, Robert and Schuh, Andreas and Webster, Matthew and Hutter, Jana and Price, Anthony and Cordero Grande, Lucilio and Hughes, Emer and Tusor, Nora and Bayly, Philip V. and Van Essen, David C. and Smith, Stephen M. and Edwards, A. David and Hajnal, Joseph and Jenkinson, Mark and Glocker, Ben and Rueckert, Daniel},
	date = {2018},
	pmid = {29100940},
	pmcid = {PMC5991912},
}

@article{robinson_msm_2014,
	title = {{MSM}: a new flexible framework for Multimodal Surface Matching},
	volume = {100},
	issn = {1095-9572},
	doi = {10.1016/j.neuroimage.2014.05.069},
	shorttitle = {{MSM}},
	pages = {414--426},
	journaltitle = {{NeuroImage}},
	shortjournal = {Neuroimage},
	author = {Robinson, Emma C. and Jbabdi, Saad and Glasser, Matthew F. and Andersson, Jesper and Burgess, Gregory C. and Harms, Michael P. and Smith, Stephen M. and Van Essen, David C. and Jenkinson, Mark},
	date = {2014-10-15},
	pmid = {24939340},
	pmcid = {PMC4190319},
}

@article{chizat_unbalanced_2019,
	title = {Unbalanced Optimal Transport: Dynamic and Kantorovich Formulation},
	url = {http://arxiv.org/abs/1508.05216},
	shorttitle = {Unbalanced Optimal Transport},
	abstract = {This article presents a new class of distances between arbitrary nonnegative Radon measures inspired by optimal transport. These distances are defined by two equivalent alternative formulations: (i) a dynamic formulation defining the distance as a geodesic distance over the space of measures (ii) a static "Kantorovich" formulation where the distance is the minimum of an optimization problem over pairs of couplings describing the transfer (transport, creation and destruction) of mass between two measures. Both formulations are convex optimization problems, and the ability to switch from one to the other depending on the targeted application is a crucial property of our models. Of particular interest is the Wasserstein-Fisher-Rao metric recently introduced independently by Chizat et al. and Kondratyev et al. Defined initially through a dynamic formulation, it belongs to this class of metrics and hence automatically benefits from a static Kantorovich formulation.},
	journaltitle = {{arXiv}:1508.05216 [math]},
	author = {Chizat, Lenaic and Peyré, Gabriel and Schmitzer, Bernhard and Vialard, François-Xavier},
	%urldate = {2021-02-26},
	date = {2019-02-09},
	eprinttype = {arxiv},
	eprint = {1508.05216},
	keywords = {Mathematics - Optimization and Control, Read, Printed},
	annotation = {Comment: 37 pages, comments welcome},
	file = {arXiv Fulltext PDF:/home/alexis/snap/zotero-snap/common/Zotero/storage/6R9EJ29I/Chizat et al. - 2019 - Unbalanced Optimal Transport Dynamic and Kantorov.pdf:application/pdf;arXiv.org Snapshot:/home/alexis/snap/zotero-snap/common/Zotero/storage/IW8MZIM3/1508.html:text/html},
}

@article{sabuncu_function-based_2010,
	title = {Function-based intersubject alignment of human cortical anatomy},
	volume = {20},
	issn = {1460-2199},
	doi = {10.1093/cercor/bhp085},
	pages = {130--140},
	number = {1},
	journaltitle = {Cerebral Cortex (New York, N.Y.: 1991)},
	shortjournal = {Cereb Cortex},
	author = {Sabuncu, Mert R. and Singer, Benjamin D. and Conroy, Bryan and Bryan, Ronald E. and Ramadge, Peter J. and Haxby, James V.},
	date = {2010-01},
	pmid = {19420007},
	pmcid = {PMC2792192},
}

@article{sejourne_unbalanced_2021,
	title = {The Unbalanced Gromov Wasserstein Distance: Conic Formulation and Relaxation},
	url = {http://arxiv.org/abs/2009.04266},
	shorttitle = {The Unbalanced Gromov Wasserstein Distance},
	abstract = {Comparing metric measure spaces (i.e. a metric space endowed with aprobability distribution) is at the heart of many machine learning problems. The most popular distance between such metric measure spaces is {theGromov}-Wasserstein ({GW}) distance, which is the solution of a quadratic assignment problem. The {GW} distance is however limited to the comparison of metric measure spaces endowed with a probability distribution.To alleviate this issue, we introduce two Unbalanced Gromov-Wasserstein formulations: a distance and a more tractable upper-bounding relaxation.They both allow the comparison of metric spaces equipped with arbitrary positive measures up to isometries. The first formulation is a positive and definite divergence based on a relaxation of the mass conservation constraint using a novel type of quadratically-homogeneous divergence. This divergence works hand in hand with the entropic regularization approach which is popular to solve large scale optimal transport problems. We show that the underlying non-convex optimization problem can be efficiently tackled using a highly parallelizable and {GPU}-friendly iterative scheme. The second formulation is a distance between mm-spaces up to isometries based on a conic lifting. Lastly, we provide numerical experiments onsynthetic examples and domain adaptation data with a Positive-Unlabeled learning task to highlight the salient features of the unbalanced divergence and its potential applications in {ML}.},
	journaltitle = {{arXiv}:2009.04266 [math, stat]},
	author = {Séjourné, Thibault and Vialard, François-Xavier and Peyré, Gabriel},
	%urldate = {2021-06-13},
	date = {2021-06-08},
	eprinttype = {arxiv},
	eprint = {2009.04266},
	keywords = {Mathematics - Optimization and Control, Statistics - Machine Learning, Read partially, Printed},
	file = {arXiv Fulltext PDF:/home/alexis/snap/zotero-snap/common/Zotero/storage/RERMXHB9/Séjourné et al. - 2021 - The Unbalanced Gromov Wasserstein Distance Conic .pdf:application/pdf;arXiv.org Snapshot:/home/alexis/snap/zotero-snap/common/Zotero/storage/WN79QZ4L/2009.html:text/html},
}

@article{memoli_gromovwasserstein_2011,
	title = {Gromov–Wasserstein Distances and the Metric Approach to Object Matching},
	volume = {11},
	issn = {1615-3383},
	url = {https://doi.org/10.1007/s10208-011-9093-5},
	doi = {10.1007/s10208-011-9093-5},
	abstract = {This paper discusses certain modifications of the ideas concerning the Gromov–Hausdorff distance which have the goal of modeling and tackling the practical problems of object matching and comparison. Objects are viewed as metric measure spaces, and based on ideas from mass transportation, a Gromov–Wasserstein type of distance between objects is defined. This reformulation yields a distance between objects which is more amenable to practical computations but retains all the desirable theoretical underpinnings. The theoretical properties of this new notion of distance are studied, and it is established that it provides a strict metric on the collection of isomorphism classes of metric measure spaces. Furthermore, the topology generated by this metric is studied, and sufficient conditions for the pre-compactness of families of metric measure spaces are identified. A second goal of this paper is to establish links to several other practical methods proposed in the literature for comparing/matching shapes in precise terms. This is done by proving explicit lower bounds for the proposed distance that involve many of the invariants previously reported by researchers. These lower bounds can be computed in polynomial time. The numerical implementations of the ideas are discussed and computational examples are presented.},
	pages = {417--487},
	number = {4},
	journaltitle = {Foundations of Computational Mathematics},
	shortjournal = {Found Comput Math},
	author = {Mémoli, Facundo},
	%urldate = {2021-06-13},
	date = {2011-08-01},
	langid = {english},
	file = {Mémoli - 2011 - Gromov–Wasserstein Distances and the Metric Approa.pdf:/home/alexis/snap/zotero-snap/common/Zotero/storage/ZC3DIDB3/Mémoli - 2011 - Gromov–Wasserstein Distances and the Metric Approa.pdf:application/pdf},
}

@article{liero_optimal_2018,
	title = {Optimal Entropy-Transport problems and a new Hellinger–Kantorovich distance between positive measures},
	volume = {211},
	issn = {1432-1297},
	url = {https://doi.org/10.1007/s00222-017-0759-8},
	doi = {10.1007/s00222-017-0759-8},
	abstract = {We develop a full theory for the new class of Optimal Entropy-Transport problems between nonnegative and finite Radon measures in general topological spaces. These problems arise quite naturally by relaxing the marginal constraints typical of Optimal Transport problems: given a pair of finite measures (with possibly different total mass), one looks for minimizers of the sum of a linear transport functional and two convex entropy functionals, which quantify in some way the deviation of the marginals of the transport plan from the assigned measures. As a powerful application of this theory, we study the particular case of Logarithmic Entropy-Transport problems and introduce the new Hellinger–Kantorovich distance between measures in metric spaces. The striking connection between these two seemingly far topics allows for a deep analysis of the geometric properties of the new geodesic distance, which lies somehow between the well-known Hellinger–Kakutani and Kantorovich–Wasserstein distances.},
	pages = {969--1117},
	number = {3},
	journaltitle = {Inventiones mathematicae},
	shortjournal = {Invent. math.},
	author = {Liero, Matthias and Mielke, Alexander and Savaré, Giuseppe},
	%urldate = {2022-05-17},
	date = {2018-03-01},
	langid = {english},
	file = {Full Text PDF:/home/alexis/snap/zotero-snap/common/Zotero/storage/WD5HDQMX/Liero et al. - 2018 - Optimal Entropy-Transport problems and a new Helli.pdf:application/pdf},
}

@inproceedings{peyre_gromov-wasserstein_2016,
	title = {Gromov-Wasserstein Averaging of Kernel and Distance Matrices},
	url = {https://proceedings.mlr.press/v48/peyre16.html},
	abstract = {This paper presents a new technique for computing the barycenter of a set of distance or kernel matrices. These matrices, which define the inter-relationships between points sampled from individual domains, are not required to have the same size or to be in row-by-row correspondence. We compare these matrices using the softassign criterion, which measures the minimum distortion induced by a probabilistic map from the rows of one similarity matrix to the rows of another; this criterion amounts to a regularized version of the Gromov-Wasserstein ({GW}) distance between metric-measure spaces. The barycenter is then defined as a Fréchet mean of the input matrices with respect to this criterion, minimizing a weighted sum of softassign values. We provide a fast iterative algorithm for the resulting nonconvex optimization problem, built upon state-of- the-art tools for regularized optimal transportation. We demonstrate its application to the computation of shape barycenters and to the prediction of energy levels from molecular configurations in quantum chemistry.},
	eventtitle = {International Conference on Machine Learning},
	pages = {2664--2672},
	booktitle = {Proceedings of The 33rd International Conference on Machine Learning},
	publisher = {{PMLR}},
	author = {Peyré, Gabriel and Cuturi, Marco and Solomon, Justin},
	%urldate = {2022-05-17},
	date = {2016-06-11},
	langid = {english},
	note = {{ISSN}: 1938-7228},
	file = {Full Text PDF:/home/alexis/snap/zotero-snap/common/Zotero/storage/ZDDNIPS8/Peyré et al. - 2016 - Gromov-Wasserstein Averaging of Kernel and Distanc.pdf:application/pdf},
}

@article{haxby_common_2011,
	title = {A common, high-dimensional model of the representational space in human ventral temporal cortex},
	volume = {72},
	issn = {1097-4199},
	doi = {10.1016/j.neuron.2011.08.026},
	pages = {404--416},
	number = {2},
	journaltitle = {Neuron},
	shortjournal = {Neuron},
	author = {Haxby, James V. and Guntupalli, J. Swaroop and Connolly, Andrew C. and Halchenko, Yaroslav O. and Conroy, Bryan R. and Gobbini, M. Ida and Hanke, Michael and Ramadge, Peter J.},
	date = {2011-10-20},
	pmid = {22017997},
	pmcid = {PMC3201764},
}

@article{vayer_fused_2018,
	title = {Fused Gromov-Wasserstein distance for structured objects: theoretical foundations and mathematical properties},
	url = {http://arxiv.org/abs/1811.02834},
	shorttitle = {Fused Gromov-Wasserstein distance for structured objects},
	abstract = {Optimal transport theory has recently found many applications in machine learning thanks to its capacity for comparing various machine learning objects considered as distributions. The Kantorovitch formulation, leading to the Wasserstein distance, focuses on the features of the elements of the objects but treat them independently, whereas the Gromov-Wasserstein distance focuses only on the relations between the elements, depicting the structure of the object, yet discarding its features. In this paper we propose to extend these distances in order to encode simultaneously both the feature and structure informations, resulting in the Fused Gromov-Wasserstein distance. We develop the mathematical framework for this novel distance, prove its metric and interpolation properties and provide a concentration result for the convergence of finite samples. We also illustrate and interpret its use in various contexts where structured objects are involved.},
	journaltitle = {{arXiv}:1811.02834 [cs, stat]},
	author = {Vayer, Titouan and Chapel, Laetita and Flamary, Rémi and Tavenard, Romain and Courty, Nicolas},
	%urldate = {2021-10-14},
	date = {2018-11-07},
	eprinttype = {arxiv},
	eprint = {1811.02834},
	keywords = {Computer Science - Machine Learning, Statistics - Machine Learning},
	file = {arXiv Fulltext PDF:/home/alexis/snap/zotero-snap/common/Zotero/storage/S8MYP7Q5/Vayer et al. - 2018 - Fused Gromov-Wasserstein distance for structured o.pdf:application/pdf;arXiv.org Snapshot:/home/alexis/snap/zotero-snap/common/Zotero/storage/G96DWWUK/1811.html:text/html},
}

@article{fischl_freesurfer_2012,
	title = {{FreeSurfer}},
	volume = {62},
	issn = {1053-8119},
	url = {https://www.sciencedirect.com/science/article/pii/S1053811912000389},
	doi = {10.1016/j.neuroimage.2012.01.021},
	series = {20 {YEARS} {OF} {fMRI}},
	abstract = {{FreeSurfer} is a suite of tools for the analysis of neuroimaging data that provides an array of algorithms to quantify the functional, connectional and structural properties of the human brain. It has evolved from a package primarily aimed at generating surface representations of the cerebral cortex into one that automatically creates models of most macroscopically visible structures in the human brain given any reasonable T1-weighted input image. It is freely available, runs on a wide variety of hardware and software platforms, and is open source.},
	pages = {774--781},
	number = {2},
	journaltitle = {{NeuroImage}},
	shortjournal = {{NeuroImage}},
	author = {Fischl, Bruce},
	%urldate = {2021-11-19},
	date = {2012-08-15},
	langid = {english},
	keywords = {Morphometry, {MRI}, Registration, Segmentation},
	file = {Accepted Version:/home/alexis/snap/zotero-snap/common/Zotero/storage/3CEGWZXM/Fischl - 2012 - FreeSurfer.pdf:application/pdf},
}

@article{yeo_spherical_2010,
	title = {Spherical Demons: Fast Diffeomorphic Landmark-Free Surface Registration},
	volume = {29},
	issn = {0278-0062},
	url = {https://www.ncbi.nlm.nih.gov/pmc/articles/PMC2862393/},
	doi = {10.1109/TMI.2009.2030797},
	shorttitle = {Spherical Demons},
	abstract = {We present the Spherical Demons algorithm for registering two spherical images. By exploiting spherical vector spline interpolation theory, we show that a large class of regularizors for the modified Demons objective function can be efficiently approximated on the sphere using iterative smoothing. Based on one parameter subgroups of diffeomorphisms, the resulting registration is diffeomorphic and fast. The Spherical Demons algorithm can also be modified to register a given spherical image to a probabilistic atlas. We demonstrate two variants of the algorithm corresponding to warping the atlas or warping the subject. Registration of a cortical surface mesh to an atlas mesh, both with more than 160k nodes requires less than 5 minutes when warping the atlas and less than 3 minutes when warping the subject on a Xeon 3.2GHz single processor machine. This is comparable to the fastest non-diffeomorphic landmark-free surface registration algorithms. Furthermore, the accuracy of our method compares favorably to the popular {FreeSurfer} registration algorithm. We validate the technique in two different applications that use registration to transfer segmentation labels onto a new image: (1) parcellation of in-vivo cortical surfaces and (2) Brodmann area localization in ex-vivo cortical surfaces.},
	pages = {650--668},
	number = {3},
	journaltitle = {{IEEE} transactions on medical imaging},
	shortjournal = {{IEEE} Trans Med Imaging},
	author = {Yeo, B.T. Thomas and Sabuncu, Mert R. and Vercauteren, Tom and Ayache, Nicholas and Fischl, Bruce and Golland, Polina},
	%urldate = {2021-11-20},
	date = {2010-03},
	pmid = {19709963},
	pmcid = {PMC2862393},
	file = {PubMed Central Full Text PDF:/home/alexis/snap/zotero-snap/common/Zotero/storage/HHWNLEJN/Yeo et al. - 2010 - Spherical Demons Fast Diffeomorphic Landmark-Free.pdf:application/pdf},
}

@article{bazeille_empirical_2021,
	title = {An empirical evaluation of functional alignment using inter-subject decoding},
	volume = {245},
	issn = {1053-8119},
	url = {https://www.sciencedirect.com/science/article/pii/S1053811921009563},
	doi = {10.1016/j.neuroimage.2021.118683},
	abstract = {Inter-individual variability in the functional organization of the brain presents a major obstacle to identifying generalizable neural coding principles. Functional alignment—a class of methods that matches subjects’ neural signals based on their functional similarity—is a promising strategy for addressing this variability. To date, however, a range of functional alignment methods have been proposed and their relative performance is still unclear. In this work, we benchmark five functional alignment methods for inter-subject decoding on four publicly available datasets. Specifically, we consider three existing methods: piecewise Procrustes, searchlight Procrustes, and piecewise Optimal Transport. We also introduce and benchmark two new extensions of functional alignment methods: piecewise Shared Response Modelling ({SRM}), and intra-subject alignment. We find that functional alignment generally improves inter-subject decoding accuracy though the best performing method depends on the research context. Specifically, {SRM} and Optimal Transport perform well at both the region-of-interest level of analysis as well as at the whole-brain scale when aggregated through a piecewise scheme. We also benchmark the computational efficiency of each of the surveyed methods, providing insight into their usability and scalability. Taking inter-subject decoding accuracy as a quantification of inter-subject similarity, our results support the use of functional alignment to improve inter-subject comparisons in the face of variable structure-function organization. We provide open implementations of all methods used.},
	pages = {118683},
	journaltitle = {{NeuroImage}},
	shortjournal = {{NeuroImage}},
	author = {Bazeille, Thomas and {DuPre}, Elizabeth and Richard, Hugo and Poline, Jean-Baptiste and Thirion, Bertrand},
	%urldate = {2021-12-16},
	date = {2021-12-15},
	langid = {english},
	keywords = {{fMRI}, Functional alignment, Inter-subject variability, Predictive modeling},
	file = {Submitted Version:/home/alexis/snap/zotero-snap/common/Zotero/storage/L7IT8WF3/Bazeille et al. - 2021 - An empirical evaluation of functional alignment us.pdf:application/pdf},
}

@article{chapel_unbalanced_2021,
	title = {Unbalanced Optimal Transport through Non-negative Penalized Linear Regression},
	url = {http://arxiv.org/abs/2106.04145},
	abstract = {This paper addresses the problem of Unbalanced Optimal Transport ({UOT}) in which the marginal conditions are relaxed (using weighted penalties in lieu of equality) and no additional regularization is enforced on the {OT} plan. In this context, we show that the corresponding optimization problem can be reformulated as a non-negative penalized linear regression problem. This reformulation allows us to propose novel algorithms inspired from inverse problems and nonnegative matrix factorization. In particular, we consider majorization-minimization which leads in our setting to efficient multiplicative updates for a variety of penalties. Furthermore, we derive for the first time an efficient algorithm to compute the regularization path of {UOT} with quadratic penalties. The proposed algorithm provides a continuity of piece-wise linear {OT} plans converging to the solution of balanced {OT} (corresponding to infinite penalty weights). We perform several numerical experiments on simulated and real data illustrating the new algorithms, and provide a detailed discussion about more sophisticated optimization tools that can further be used to solve {OT} problems thanks to our reformulation.},
	journaltitle = {{arXiv}:2106.04145 [cs, math, stat]},
	author = {Chapel, Laetitia and Flamary, Rémi and Wu, Haoran and Févotte, Cédric and Gasso, Gilles},
	%urldate = {2022-02-23},
	date = {2021-06-08},
	eprinttype = {arxiv},
	eprint = {2106.04145},
	keywords = {Computer Science - Machine Learning, Mathematics - Optimization and Control, Statistics - Machine Learning},
	annotation = {Comment: Laetitia Chapel and R{\textbackslash}'emi Flamary have equal contribution},
	file = {arXiv Fulltext PDF:/home/alexis/snap/zotero-snap/common/Zotero/storage/EYAHFKFS/Chapel et al. - 2021 - Unbalanced Optimal Transport through Non-negative .pdf:application/pdf;arXiv.org Snapshot:/home/alexis/snap/zotero-snap/common/Zotero/storage/KYRTJV8P/2106.html:text/html},
}

@misc{gramfort2015,
  doi = {10.48550/ARXIV.1503.08596},
  url = {https://arxiv.org/abs/1503.08596},
  author = {Gramfort, Alexandre and Peyré, Gabriel and Cuturi, Marco},
  title = {Fast Optimal Transport Averaging of Neuroimaging Data},
  publisher = {arXiv},
  year = {2015},  
  copyright = {arXiv.org perpetual, non-exclusive license}
}

@ARTICLE{Glasser2016-ha,
  title     = "A multi-modal parcellation of human cerebral cortex",
  author    = "Glasser, Matthew F and Coalson, Timothy S and Robinson, Emma C
               and Hacker, Carl D and Harwell, John and Yacoub, Essa and
               Ugurbil, Kamil and Andersson, Jesper and Beckmann, Christian F
               and Jenkinson, Mark and Smith, Stephen M and Van Essen, David C",
  journal   = "Nature",
  publisher = "nature.com",
  volume    =  536,
  number    =  7615,
  pages     = "171--178",
  month     =  aug,
  year      =  2016,
  language  = "en"
}

@article{ants,
    author = {Avants, B.B. and Epstein, C.L. and Grossman, M. and Gee, J.C.},
    doi = {10.1016/j.media.2007.06.004},
    issn = {1361-8415},
    journal = {Medical Image Analysis},
    number = 1,
    pages = {26-41},
    shorttitle = {Symmetric diffeomorphic image registration with cross-correlation},
    title = {Symmetric diffeomorphic image registration with cross-correlation: Evaluating automated labeling of elderly and neurodegenerative brain},
    url = {http://www.sciencedirect.com/science/article/pii/S1361841507000606},
    volume = 12,
    year = 2008
}

@article{fs_reconall,
    author = {Dale, Anders M. and Fischl, Bruce and Sereno, Martin I.},
    doi = {10.1006/nimg.1998.0395},
    issn = {1053-8119},
    number = 2,
    pages = {179-194},
    shorttitle = {Cortical Surface-Based Analysis},
    title = {Cortical Surface-Based Analysis: I. Segmentation and Surface Reconstruction},
    url = {http://www.sciencedirect.com/science/article/pii/S1053811998903950},
    volume = 9,
    year = 1999
}

@Article{pizzigali2020,
   Author="Pizzagalli, F.  and Auzias, G.  and Yang, Q.  and Mathias, S. R.  and Faskowitz, J.  and Boyd, J. D.  and Amini, A.  and Rivière, D.  and McMahon, K. L.  and de Zubicaray, G. I.  and Martin, N. G.  and Mangin, J. F.  and Glahn, D. C.  and Blangero, J.  and Wright, M. J.  and Thompson, P. M.  and Kochunov, P.  and Jahanshad, N. ",
   Title="{{T}he reliability and heritability of cortical folds and their genetic correlations across hemispheres}",
   Journal="Commun Biol",
   Year="2020",
   Volume="3",
   Number="1",
   Pages="510",
   Month="09"
}

@Article{vanessen2012,
   Author="Van Essen, D. C.  and Glasser, M. F.  and Dierker, D. L.  and Harwell, J.  and Coalson, T. ",
   Title="{{P}arcellations and hemispheric asymmetries of human cerebral cortex analyzed on surface-based atlases}",
   Journal="Cereb Cortex",
   Year="2012",
   Volume="22",
   Number="10",
   Pages="2241--2262",
   Month="10"
}

@Article{alwasity2020,
   Author="Al-Wasity, S.  and Vogt, S.  and Vuckovic, A.  and Pollick, F. E. ",
   Title="{{H}yperalignment of motor cortical areas based on motor imagery during action observation}",
   Journal="Sci Rep",
   Year="2020",
   Volume="10",
   Number="1",
   Pages="5362",
   Month="03"
}

@inproceedings{Chen2015,
 author = {Chen, Po-Hsuan (Cameron) and Chen, Janice and Yeshurun, Yaara and Hasson, Uri and Haxby, James and Ramadge, Peter J},
 booktitle = {Advances in Neural Information Processing Systems},
 editor = {C. Cortes and N. Lawrence and D. Lee and M. Sugiyama and R. Garnett},
 pages = {},
 publisher = {Curran Associates, Inc.},
 title = {A Reduced-Dimension fMRI Shared Response Model},
 url = {https://proceedings.neurips.cc/paper/2015/file/b3967a0e938dc2a6340e258630febd5a-Paper.pdf},
 volume = {28},
 year = {2015}
}

@article{ibc,
  TITLE = {{Individual Brain Charting, a high-resolution fMRI dataset for cognitive mapping}},
  AUTHOR = {Pinho, Ana Lu{\'i}sa and Amadon, Alexis and Ruest, Torsten and Fabre, Murielle and Dohmatob, Elvis and Denghien, Isabelle and Ginisty, Chantal and Becuwe-Desmidt, S{\'e}verine and Roger, S{\'e}verine and Laurier, Laurence and Joly-Testault, V{\'e}ronique and M{\'e}diouni-Cloarec, Ga{\"e}lle and Doubl{\'e}, Christine and Martins, Bernadette and Pinel, Philippe and Eger, Evelyn and Varoquaux, Gael and Pallier, Christophe and Dehaene, Stanislas and Hertz-Pannier, Lucie and Thirion, Bertrand},
  URL = {https://hal.archives-ouvertes.fr/hal-01817528},
  JOURNAL = {{Scientific Data }},
  PUBLISHER = {{Nature Publishing Group}},
  VOLUME = {5},
  PAGES = {180105},
  YEAR = {2018},
  MONTH = Jun,
  DOI = {10.1038/sdata.2018.105},
  HAL_ID = {hal-01817528},
  HAL_VERSION = {v1},
}

@article{thirion:2014,
  TITLE = {{Which fMRI clustering gives good brain parcellations?}},
  AUTHOR = {Thirion, Bertrand and Varoquaux, Ga{\"e}l and Dohmatob, Elvis and Poline, Jean-Baptiste},
  URL = {https://hal.inria.fr/hal-01015172},
  JOURNAL = {{Frontiers in Neuroscience}},
  PUBLISHER = {{Frontiers}},
  VOLUME = {8},
  NUMBER = {167},
  PAGES = {13},
  YEAR = {2014},
  MONTH = May,
  DOI = {10.3389/fnins.2014.00167},
  HAL_ID = {hal-01015172},
  HAL_VERSION = {v1},
}

@book{memoli_use_2007,
	title = {On the use of Gromov-Hausdorff Distances for Shape Comparison},
	isbn = {978-3-905673-51-7},
	url = {https://diglib.eg.org:443/xmlui/handle/10.2312/SPBG.SPBG07.081-090},
	abstract = {It is the purpose of this paper to propose and discuss certain modifications of the ideas concerning Gromov- Hausdorff distances in order to tackle the problems of shape matching and comparison. These reformulations render these distances more amenable to practical computations without sacrificing theoretical underpinnings. A second goal of this paper is to establish links to several other practical methods proposed in the literature for comparing/matching shapes in precise terms. Connections with the Quadratic Assignment Problem ({QAP}) are also established, and computational examples are presented.},
	publisher = {The Eurographics Association},
	author = {Memoli, Facundo},
	%urldate = {2022-05-18},
	date = {2007},
	langid = {english},
	doi = {10.2312/SPBG/SPBG07/081-090},
	note = {Accepted: 2014-01-29T16:52:11Z
{ISSN}: 1811-7813},
	file = {Full Text PDF:/home/alexis/snap/zotero-snap/common/Zotero/storage/N4DN3MMV/Memoli - 2007 - On the use of Gromov-Hausdorff Distances for Shape.pdf:application/pdf;Snapshot:/home/alexis/snap/zotero-snap/common/Zotero/storage/XCBGRJX4/SPBG.SPBG07.html:text/html},
}

@Article{hcpdata,
   Author="Van Essen, D. C.  and Smith, S. M.  and Barch, D. M.  and Behrens, T. E.  and Yacoub, E.  and Ugurbil, K.  and Van Essen, D.  and Barch, D.  and Corbetta, M.  and Goate, A.  and Heath, A.  and Larson-Prior, L.  and Marcus, D.  and Petersen, S.  and Prior, F.  and Province, M.  and Raichle, M.  and Schlaggar, B.  and Shimony, J.  and Snyder, A.  and Adeyemo, B.  and Archie, K.  and Babajani-Feremi, A.  and Bloom, N.  and Bryant, J. E.  and Burgess, G.  and Cler, E.  and Coalson, T.  and Curtiss, S.  and Danker, S.  and Denness, R.  and Dierker, D.  and Elam, J.  and Evans, T.  and Feldt, C.  and Fenlon, K.  and Footer, O.  and Glasser, M.  and Gordon, E.  and Gu, P.  and Guilday, C.  and Harms, M.  and Hartley, T.  and Harwell, J.  and Hileman, M.  and Hodge, M.  and Hood, L.  and Horton, W.  and House, M.  and Laumann, T.  and Lugo, M.  and Marion, S.  and Miezin, F.  and Nolan, D.  and Nolan, T.  and Power, J.  and Ramaratnam, M.  and Reid, E.  and Schindler, J.  and Schmitz, D.  and Schweiss, C.  and Serati, J.  and Taylor, B.  and Tobias, M.  and Wilson, T.  and Ugurbil, K.  and Garwood, M.  and Harel, N.  and Lenglet, C.  and Yacoub, E.  and Adriany, G.  and Auerbach, E.  and Moeller, S.  and Strupp, J.  and Smith, S.  and Behrens, T.  and Jenkinson, M.  and Johansen-Berg, H.  and Miller, K.  and Woolrich, M.  and Andersson, J.  and Duff, E.  and Hernandez, M.  and Jbabdi, S.  and Robinson, E.  and Salimi-Khorshidi, R.  and Sotiropoulos, S.  and Romani, G. L.  and Della Penna, S.  and Pizzella, V.  and de Pasquale, F.  and Di Pompeo, F.  and Marzetti, L.  and Perruci, G.  and Bucholz, R.  and Roskos, T.  and Kiser, T.  and Luo, Q. J.  and Stout, J.  and Oostenveld, R.  and Beckmann, C.  and Schoffelen, J. M.  and Fries, P.  and Michalareas, G.  and Sapiro, G.  and Sporns, O.  and Nichols, T.  and Farber, G.  and Bjork, J.  and Blumensath, T.  and Chang, A.  and Chen, L.  and Feinberg, D.  and Kull, L.  and Wig, G.  and Xu, J. G.  and Basser, P.  and Bullmore, E.  and Evans, A.  and Gazzaniga, M.  and Glahn, D.  and Hawrylycz, M.  and Hennig, J.  and Parker, G.  and Poldrack, R.  and Salmelin, R. ",
   Title="{{T}he {W}{U}-{M}inn {H}uman {C}onnectome {P}roject: an overview}",
   Journal="Neuroimage",
   Year="2013",
   Volume="80",
   Pages="62--79",
   Month="10"
}

@article{Tavor2016-rl,
  title={Task-free MRI predicts individual differences in brain activity during task performance},
  author={Tavor, Ido and Jones, O Parker and Mars, Rogier B and Smith, SM and Behrens, TE and Jbabdi, Saad},
  journal={Science},
  volume={352},
  number={6282},
  pages={216--220},
  year={2016},
  publisher={American Association for the Advancement of Science}
}

@incollection{NEURIPS2019_9015,
title = {PyTorch: An Imperative Style, High-Performance Deep Learning Library},
author = {Paszke, Adam and Gross, Sam and Massa, Francisco and Lerer, Adam and Bradbury, James and Chanan, Gregory and Killeen, Trevor and Lin, Zeming and Gimelshein, Natalia and Antiga, Luca and Desmaison, Alban and Kopf, Andreas and Yang, Edward and DeVito, Zachary and Raison, Martin and Tejani, Alykhan and Chilamkurthy, Sasank and Steiner, Benoit and Fang, Lu and Bai, Junjie and Chintala, Soumith},
booktitle = {Advances in Neural Information Processing Systems 32},
editor = {H. Wallach and H. Larochelle and A. Beygelzimer and F. d\textquotesingle Alch\'{e}-Buc and E. Fox and R. Garnett},
pages = {8024--8035},
year = {2019},
publisher = {Curran Associates, Inc.},
url = {http://papers.neurips.cc/paper/9015-pytorch-an-imperative-style-high-performance-deep-learning-library.pdf}
}

@article{scikit-learn,
 title={Scikit-learn: Machine Learning in {P}ython},
 author={Pedregosa, F. and Varoquaux, G. and Gramfort, A. and Michel, V.
         and Thirion, B. and Grisel, O. and Blondel, M. and Prettenhofer, P.
         and Weiss, R. and Dubourg, V. and Vanderplas, J. and Passos, A. and
         Cournapeau, D. and Brucher, M. and Perrot, M. and Duchesnay, E.},
 journal={Journal of Machine Learning Research},
 volume={12},
 pages={2825--2830},
 year={2011}
}

@article{abraham_machine_2014,
	title = {Machine learning for neuroimaging with scikit-learn},
	volume = {8},
	issn = {1662-5196},
	url = {https://www.frontiersin.org/article/10.3389/fninf.2014.00014},
	journaltitle = {Frontiers in Neuroinformatics},
	author = {Abraham, Alexandre and Pedregosa, Fabian and Eickenberg, Michael and Gervais, Philippe and Mueller, Andreas and Kossaifi, Jean and Gramfort, Alexandre and Thirion, Bertrand and Varoquaux, Gael},
	urldate = {2022-05-19},
	date = {2014},
}

@article{schneider2019,
author = {Marian Schneider  and Valentin G. Kemper  and Thomas C. Emmerling  and Federico De Martino  and Rainer Goebel },
title = {Columnar clusters in the human motion complex reflect consciously perceived motion axis},
journal = {Proceedings of the National Academy of Sciences},
volume = {116},
number = {11},
pages = {5096-5101},
year = {2019},
doi = {10.1073/pnas.1814504116},
URL = {https://www.pnas.org/doi/abs/10.1073/pnas.1814504116},
eprint = {https://www.pnas.org/doi/pdf/10.1073/pnas.1814504116}}

@article{richard2020modeling,
  title={Modeling shared responses in neuroimaging studies through multiview ica},
  author={Richard, Hugo and Gresele, Luigi and Hyvarinen, Aapo and Thirion, Bertrand and Gramfort, Alexandre and Ablin, Pierre},
  journal={Advances in Neural Information Processing Systems},
  volume={33},
  pages={19149--19162},
  year={2020}
}

@article{nishimoto2011reconstructing,
  title={Reconstructing visual experiences from brain activity evoked by natural movies},
  author={Nishimoto, Shinji and Vu, An T and Naselaris, Thomas and Benjamini, Yuval and Yu, Bin and Gallant, Jack L},
  journal={Current biology},
  volume={21},
  number={19},
  pages={1641--1646},
  year={2011},
  publisher={Elsevier}
}

@article{bhattasali2019localising,
  title={Localising memory retrieval and syntactic composition: an fMRI study of naturalistic language comprehension},
  author={Bhattasali, Shohini and Fabre, Murielle and Luh, Wen-Ming and Al Saied, Hazem and Constant, Mathieu and Pallier, Christophe and Brennan, Jonathan R and Spreng, R Nathan and Hale, John},
  journal={Language, Cognition and Neuroscience},
  volume={34},
  number={4},
  pages={491--510},
  year={2019},
  publisher={Taylor \& Francis}
}

@article{destrieux_automatic_2010,
	title = {Automatic parcellation of human cortical gyri and sulci using standard anatomical nomenclature},
	volume = {53},
	issn = {1053-8119},
	url = {https://www.sciencedirect.com/science/article/pii/S1053811910008542},
	doi = {10.1016/j.neuroimage.2010.06.010},
	pages = {1--15},
	number = {1},
	journaltitle = {{NeuroImage}},
	shortjournal = {{NeuroImage}},
	author = {Destrieux, Christophe and Fischl, Bruce and Dale, Anders and Halgren, Eric},
	urldate = {2022-08-02},
	date = {2010-10-15},
	langid = {english},
}

\end{document}